\documentclass[acmsmall,authorversion=true]{acmart}

\AtBeginDocument{%
  \providecommand\BibTeX{{%
    \normalfont B\kern-0.5em{\scshape i\kern-0.25em b}\kern-0.8em\TeX}}}

\setcopyright{acmcopyright}
\acmDOI{10.1145/3491244}

\received{February 2021}
\received[revised]{September 2021}
\received[accepted]{October 2021}

\acmJournal{TIOT}
\acmYear{2022} 
\acmVolume{3} 
\acmNumber{2} 
\acmArticle{11} 
\acmMonth{2} 
\acmPrice{15.00}

\usepackage{booktabs} 
\usepackage[acronym, nowarn]{glossaries}
\makeglossaries

\usepackage{amsfonts}
\usepackage{amsthm}
\usepackage{amsmath}
\usepackage{algorithmic}
\usepackage[ruled, vlined]{algorithm2e}
\usepackage{array}
\usepackage{arydshln}

\usepackage{graphicx} 

\usepackage{xcolor}
\usepackage{cryptocode}
\usepackage{caption}
\usepackage{subcaption}
\usepackage{xspace}
\usepackage{listings}
\usepackage{pgfplots,pgfplotstable}
\usepackage{tikz}
\usetikzlibrary{matrix,decorations.pathreplacing, calc, chains, positioning}
\usepackage{pythonhighlight}
\usepackage[export]{adjustbox}
\usepackage[normalem]{ulem}

\usepackage{hyperref}
\usepackage{cleveref}

\usepackage{enumitem}

\usepackage{xargs}   
\usepackage{xcolor,colortbl}  

\definecolor{Gray}{gray}{0.8}
\definecolor{OliveGreen}{rgb}{0,0.6,0}
\definecolor{Purple}{rgb}{0.49, 0.45, 0.71}
\definecolor{AntiqueBrass}{rgb}{0.8, 0.58, 0.46}

\definecolor{path0}{rgb}{1, 0, 0}
\definecolor{path1}{rgb}{0.9412, 0.6235, 0.0549}
\definecolor{path2}{rgb}{0.5725, 0.8157, 0.3137}
\definecolor{path3}{rgb}{0.3569, 0.6078, 0.8353}

\definecolor{mygreen}{HTML}{15b01a}
\definecolor{myred}{HTML}{e50000}
\definecolor{mytangerine}{HTML}{ff9408}
\definecolor{mysky}{HTML}{448ee4}

\newcommand\undermat[2]{%
\makebox[0pt][l]{$\smash{\underbrace{\phantom{%
\begin{matrix}#2\end{matrix}}}_{\text{$#1$}}}$}#2}

\makeatletter
\newcommand*\bigcdot{\mathpalette\bigcdot@{.7}}
\newcommand*\bigcdot@[2]{\mathbin{\vcenter{\hbox{\scalebox{#2}{$\m@th#1\bullet$}}}}}
\makeatother

\usepackage[colorinlistoftodos,prependcaption,textsize=tiny]{todonotes}
\newcommandx{\unsure}[2][1=]{\todo[linecolor=red,backgroundcolor=red!25,bordercolor=red,#1]{#2}}
\newcommandx{\change}[2][1=]{\todo[linecolor=blue,backgroundcolor=blue!25,bordercolor=blue,#1]{#2}}
\newcommandx{\info}[2][1=]{\todo[linecolor=OliveGreen,backgroundcolor=OliveGreen!25,bordercolor=OliveGreen,#1]{#2}}
\newcommandx{\improvement}[2][1=]{\todo[linecolor=Plum,backgroundcolor=Plum!25,bordercolor=Plum,#1]{#2}}
\newcommandx{\thiswillnotshow}[2][1=]{\todo[disable,#1]{#2}}

\newcommand{\name}{\textit{Next2You}\xspace}

\usepackage{newtxmath}

\usepackage{mathtools}

\usepackage{soul}

\usepackage{makecell}

\usepackage{multirow}

\usepackage{nicematrix}
\usetikzlibrary{decorations.pathreplacing}

\usepackage{tikz}
\usetikzlibrary{arrows.meta}
\usepackage[edges]{forest}
\usetikzlibrary{shapes.geometric}
\usetikzlibrary{backgrounds}
\definecolor{mygreen}{RGB}{0,173,67}
\usepackage{transparent}

\usepackage{textcomp}
\usepackage{siunitx}

\definecolor{mygreen}{HTML}{15b01a}
\definecolor{myred}{HTML}{e50000}

\forestset{
  declare dimen register=gap,
  gap'=4mm,
  declare count register=twist,
  twist'=1,
  family tree too/.style={
    draw,
    for tree={
      fill=gray!0,
      rounded corners=1pt,
      edge=thick,
    },
    where={level()<(twist)}{%
      parent anchor=children,
    }{%
      draw,
      folder,
      grow'=0,
      if={level()==(twist)}{%
        before typesetting nodes={child anchor=north},
        anchor=north,
        edge path'={(!u.parent anchor) -- ++(0,-4pt) -| (.child anchor)},
      }{
        if level=2{
          l sep = 4mm,
                    before computing xy={
                      l/.wrap pgfmath arg={##1}{10pt-((\textwidth-6*(gap))/6)}
                    },
        }{}
      },
    },
  },
}

\newacronym{ap}{AP}{access point}
\newacronym{aoa}{AoA}{angle of arrival}
\newacronym{ble}{BLE}{Bluetooth Low Energy}
\newacronym{iot}{\textsc{IoT}}{Internet of Things}
\newacronym{zip}{\textsc{ZIP}}{zero-interaction pairing}
\newacronym{zia}{\textsc{ZIA}}{zero-interaction authentication}
\newacronym{zis}{\textsc{ZIS}}{zero-interaction security}
\newacronym{far}{FAR}{False Acceptance Rate}
\newacronym{frr}{FRR}{False Rejection Rate}
\newacronym{eer}{EER}{Equal Error Rate}
\newacronym{pki}{PKI}{Public Key Infrastructure}
\newacronym{fft}{FFT}{fast Fourier transform}
\newacronym{ntp}{NTP}{Network Time Protocol}
\newacronym{imu}{IMU}{inertial measurement unit}
\newacronym{csi}{CSI}{channel state information}
\newacronym{cir}{CIR}{channel impulse response}
\newacronym{cv}{CV}{cross-validation}
\newacronym{auc}{AUC}{Area Under the Curve}
\newacronym{qos}{QoS}{Quality of service}
\newacronym{agc}{AGC}{Automatic gain control}
\newacronym{rrr}{RRR}{Right for the Right Reasons}
\newacronym{flops}{FLOPS}{floating point operations per second}
\newacronym{snr}{SNR}{signal-to-noise ratio}
\newacronym{pdp}{PDP}{power delay profile}
\newacronym{tof}{ToF}{time of flight}
\newacronym{rss}{RSS}{received signal strength}

\newacronym{toa}{ToA}{time of arrival}
\newacronym{tdoa}{TDoA}{time difference of arrival}

\begin{document}

\title{Next2You: Robust Copresence Detection Based on Channel State Information}

\author{Mikhail Fomichev}
\orcid{0000-0001-9697-0359}
\affiliation{%
 \department{Secure Mobile Networking Lab}
 \institution{Technical University of Darmstadt}
 \country{Germany}}
\email{mfomichev@seemoo.tu-darmstadt.de}

\author{Luis F. Abanto-Leon}
\orcid{0000-0002-3533-7928}        
\affiliation{%
 \department{Secure Mobile Networking Lab}
 \institution{Technical University of Darmstadt}
 \country{Germany}}
\email{labanto@seemoo.tu-darmstadt.de}

\author{Max Stiegler}
\orcid{0000-0001-8420-7217}
\affiliation{%
 \department{Secure Mobile Networking Lab}
 \institution{Technical University of Darmstadt}
 \country{Germany}}
\email{mstiegler@seemoo.tu-darmstadt.de}

\author{Alejandro Molina}
\orcid{0000-0003-4509-9174}
\affiliation{%
 \department{Machine Learning Group}
 \institution{Technical University of Darmstadt}
 \country{Germany}}
\email{molina@cs.tu-darmstadt.de}

\author{Jakob Link}
\orcid{0000-0003-1068-6053}        
\affiliation{%
 \department{Secure Mobile Networking Lab}
 \institution{Technical University of Darmstadt}
 \country{Germany}}
\email{jlink@seemoo.tu-darmstadt.de}

\author{Matthias Hollick}
\orcid{0000-0002-9163-5989}
\affiliation{%
 \department{Secure Mobile Networking Lab}
 \institution{Technical University of Darmstadt}
 \country{Germany}}
\email{mhollick@seemoo.tu-darmstadt.de}

\renewcommand{\shortauthors}{M. Fomichev et al.}

\begin{abstract}

Context-based copresence detection schemes are a necessary prerequisite to building secure and usable authentication systems in the \gls{iot}. Such schemes allow one device to verify proximity of another device without user assistance utilizing their physical context (e.g., audio). 
The state-of-the-art copresence detection schemes suffer from two major limitations: (1) they cannot accurately detect copresence in low-entropy context (e.g., empty room with few events occurring) and insufficiently separated environments (e.g., adjacent rooms), (2) they require devices to have common sensors (e.g., microphones) to capture context, making them impractical on devices with heterogeneous sensors.
We address these limitations, proposing \name, a novel copresence detection scheme utilizing \gls{csi}.
In particular, we leverage magnitude and phase values from a range of subcarriers specifying a Wi-Fi channel to capture a robust wireless context created when devices communicate.  
We implement \name on off-the-shelf smartphones relying only on ubiquitous Wi-Fi chipsets and evaluate it based on over 95 hours of \gls{csi} measurements that we collect in five real-world scenarios. 
\name achieves error rates below 4\%, maintaining accurate copresence detection both in low-entropy context and insufficiently separated environments.
We also demonstrate the capability of \name to work reliably in real-time and its robustness to various attacks.

\end{abstract}



\maketitle


\glsresetall

\section{Introduction}
\label{sec:intro}
The proliferation of the \gls{iot} urges the need for authentication systems that are both secure and usable.  
Copresence detection is a necessary prerequisite to \gls{zia}---a technique that allows one device to authenticate another based on their physical proximity. 
Specifically, copresence detection is used to mitigate relay attacks~\cite{Truong:2014, Shrestha:2014, Truong:2019} or serve as a second authentication factor~\cite{Marforio:2014, Mare:2014, Karapanos:2015}, providing improved \textit{usability} by minimizing user interaction. 
To achieve this, copresence detection schemes utilize devices' \textit{context}, which is represented as a set of \textit{sensor modalities} (e.g., audio, signal strength) collected by devices from their ambient environment, hence they are known as \textit{context-based}~\cite{Truong:2019, Han:2018, Fomichev:2019}.  
In addition to usability, context-based schemes have two extra advantages compared to user-assisted copresence detection methods (e.g., entering a verification code)~\cite{Han:2018, Fomichev:2019}. 
First, they scale better to an ever-growing number of devices, including those that lack user interfaces.  
Second, context-based schemes rely on existing sensing capabilities of \gls{iot} devices, without requiring any established infrastructure (e.g., PKI) or dedicated hardware (e.g., NFC), facilitating interoperability~\cite{Fomichev:2019, Karapanos:2015}. 

The rationale for context-based copresence detection is that \textit{copresent} devices, located inside an enclosed physical space (e.g., a room), will perceive similar context compared to devices outside this space. 
Hence, the security of such schemes relies on the unpredictability of the shared context, which depends on the intensity and variety of ambient activity such as sound or motion happening in the environment. 
To date, a number of context-based copresence detection schemes utilizing different sensor modalities have been proposed~\cite{Halevi:2012, Truong:2014, Mare:2014, Shrestha:2014, Karapanos:2015, Shepherd:2017, Truong:2019, Shrestha:2019}.
Some schemes are already used in commercial products such as Futurae Authentication~\cite{Futurae:2020} and Apple Auto Unlock~\cite{Stute:2018}, making their security and utility crucial for real-world applications.  
However, these state-of-the-art schemes have two major limitations. 
First, they show reduced copresence detection accuracy, hence lower security, in the cases of \textit{low-entropy context} (e.g., empty room with few events occurring) and \textit{insufficiently separated environments} (e.g., adjacent rooms). In the first case, the context becomes predictable allowing the adversary to guess or manipulate it in a controlled manner~\cite{Fomichev:2019, Shrestha:2016spf}. 
In the second case, close environments partly share the context (e.g., loud sound), confusing copresence detection if several devices start it simultaneously~\cite{Stute:2019, Fomichev:2019}.  
Second, the state-of-the-art context-based schemes require devices to be equipped with common sensors such as microphones, limiting their utility, because many \gls{iot} devices have only a single sensor (e.g., power meter)~\cite{Han:2018}. 

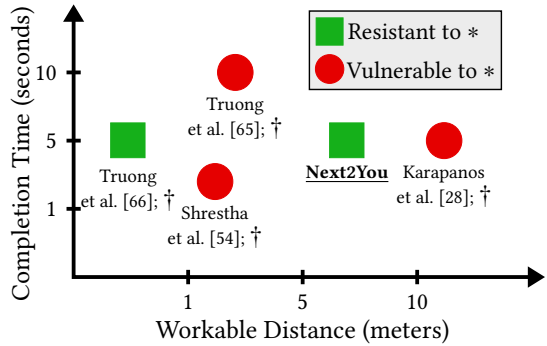
\begin{figure}
\centering
\resizebox{0.532\textwidth}{!}{%
	\begin{tikzpicture}[square/.style={regular polygon,regular polygon sides=4}]
		\draw [>=Triangle, <->, ultra thick] (0,4) node (yaxis) {} |- (7,0) node (xaxis) {};
		
		\draw [ultra thick] (3pt,1) -- (-3pt,1) node[anchor=east] {\normalsize 1}; 
		\draw [ultra thick] (3pt,2) -- (-3pt,2) node[anchor=east] {\normalsize 5};
		\draw [ultra thick] (3pt,3) -- (-3pt,3) node[anchor=east] {\normalsize 10};
		
		\draw [ultra thick] (1.7,3pt) -- (1.7,-3pt) node[anchor=north] {\normalsize 1};
		\draw [ultra thick] (3.4,3pt) -- (3.4,-3pt) node[anchor=north] {\normalsize 5};
		\draw [ultra thick] (5.1,3pt) -- (5.1,-3pt) node[anchor=north] {\normalsize 10};
		
		\draw [mygreen, fill=mygreen, text=black] (3.8, 1.75) rectangle ++(0.51cm, 0.51cm) 
		node[xshift=-0.25cm, yshift=-0.5cm, anchor=north] {\footnotesize \underline{\textbf{Next2You}}}; 
		
		\draw [mygreen, fill=mygreen, text=black] (0.55, 1.75) rectangle ++(0.51cm, 0.51cm) 
		node[xshift=-0.25cm, yshift=-0.5cm, anchor=north] 
		{\footnotesize \begin{tabular}{c} Truong \\ et al. \cite{Truong:2019}; \large $\dagger$ \end{tabular}};
		
		\draw[myred,fill=myred, text=black] (2.1, 1.4) circle (7.5 pt) node[yshift=-0.18cm, anchor=north] 
		{\footnotesize \begin{tabular}{c} Shrestha \\ et al. \cite{Shrestha:2014}; \large $\dagger$ \end{tabular}};
		
		\draw[myred,fill=myred, text=black] (2.4, 3) circle (7.5 pt) node[yshift=-0.18cm, anchor=north] 
		{\footnotesize \begin{tabular}{c} Truong \\ et al. \cite{Truong:2014}; \large $\dagger$ \end{tabular}};
		
		\draw[myred,fill=myred, text=black] (5.5, 2) circle (7.5 pt) node[yshift=-0.18cm, anchor=north] 
		{\footnotesize \begin{tabular}{c} Karapanos \\ et al. \cite{Karapanos:2015}; \large $\dagger$ \end{tabular}};
		
		\draw [black, semithick, fill=gray!15, show background rectangle] (3.5, 3.9) rectangle ++(2.85, -1.15);
		
		\draw [mygreen, fill=mygreen, text=black] (3.6, 3.4) rectangle ++(0.4cm, 0.4cm) 
		node[xshift=-0.05cm, yshift=-0.19cm, anchor=west] {\normalsize Resistant to \large $*$};
		
		\draw[myred,fill=myred, text=black] (3.8, 3.05) circle (6pt)
		node[xshift=0.13cm, yshift=-0.02cm, anchor=west] {\normalsize Vulnerable to \large $*$};
		
		\node[xshift=-3.6cm, below=0.5cm] at (xaxis) {\large Workable Distance (meters)};
		\node[yshift=0.09cm, rotate=90, left=0.75cm] at (yaxis) {\large Completion Time (seconds)};
	\end{tikzpicture}
}%
	\caption{Design space of \name: comparison with state-of-the-art schemes. \name prevents attacks caused by low-entropy context and insufficiently separated environments ({\large $*$}), and it \textit{does not} require devices to have common sensors ({\large $\dagger$}). Similar to the state of the art, \name provides reliable copresence detection at distances of several meters, requiring a few seconds to complete the procedure.}
	\label{fig:n2u-design}
\end{figure}

The above two limitations impair security and utility of context-based copresence detection schemes, hindering their adoption in the \gls{iot}. 
We address these limitations proposing \name, a novel copresence detection scheme based on \gls{csi}.  
\autoref{fig:n2u-design} shows the design space of \name in comparison to state-of-the-art schemes. 
Specifically, \name has higher security, achieving accurate copresence detection in low-entropy context and insufficiently separated environments, and is deployable on devices with heterogeneous sensors, while performing similarly in terms of completion time and distance at which copresence detection is viable.   

To the best of our knowledge, we are the first to demonstrate the feasibility of copresence detection based on \gls{csi}, leveraging its two advantages. 
First, \gls{csi} is mandatory information generated when Wi-Fi enabled devices communicate.  
The ubiquity of Wi-Fi in \gls{iot} devices~\cite{WiFi:2019} and increasing \gls{csi} availability in them~\cite{Halperin:2011, Xie:2015, Nexmon:2017} allow \name to run on devices that do not have common sensors but are equipped with Wi-Fi (e.g., a laptop and smart plug). 
Second, \gls{csi} is known to be location-sensitive, capturing variation of a wireless channel, which is affected by distance between devices, geometry and materials of surroundings (e.g., walls) as well as wireless spectrum busyness and mobility~\cite{Xi:2016, Ma:2019}. Prior work has extensively utilized \gls{csi} for indoor device localization, including security-focused solutions~\cite{Xiong:2013, Fang:2014, Fang:2016}. 
In~\autoref{sec:back-work}, we review indoor localization schemes, showing their fundamental differences with \name in terms of purpose and operation (e.g., infrastructure- vs. ad hoc-based). 
Also, the assumptions made by many of these schemes are unrealistic for the \gls{iot}, namely they presume that devices have multiple antennas, significant processing power, and tight time synchronization, being able to capture hundreds of \gls{csi} measurements per second.
Our goal is different, as we aim to detect copresence (e.g., if devices are in the same room) and not precise location.
This is challenging because we need to obtain common \gls{csi} features for copresent devices within the same environment, ensuring that such features are distinctive among non-copresent devices in different environments.
In addition, we evaluate \name on smartphones and the Raspberry Pi which have hardware typical of \gls{iot} devices (e.g., portable, one antenna), assuming realistic conditions such as loose time synchronization between devices and a few \gls{csi} measurements per second. 
Apart from \gls{zia}, \name can enhance physical layer security by enabling a covert channel between colocated devices~\cite{Schulz:2018} or be adapted for user or object occupancy detection~\cite{Gringoli:2019, Zhu:2020}.

To achieve our goal, we propose using \gls{csi} magnitude and phase values in a neural network to capture robust wireless context commonly observed by copresent devices. 
We design the network such that it not only automatically learns relevant copresence features, which we corroborate with prior research, but also enables generalizability and high performance of \name. 
To demonstrate the effectiveness of \name, we collect \gls{csi} data in five real-world scenarios, including a busy office, an urban apartment, a rural house as well as parked and moving cars, resulting in over 95 hours of \gls{csi} measurements.
We show that \name provides reliable copresence detection with error rates below 4\% in both 2.4 GHz and 5 GHz frequency bands, and it is capable of running on off-the-shelf smartphones in real-time.  
\name maintains accurate copresence detection in challenging cases of low-entropy context and insufficiently separated environments.
Through our real-world experiments, we demonstrate the robustness of \name copresence detection, its ability to generalize to new application scenarios, and its resilience to attacks. 

In summary, we make the following contributions:
\begin{itemize}
	\item We design \name, a novel copresence detection scheme that combines \gls{csi} and neural networks, justifying their mutual suitability for copresence detection.
	\item We collect a real-world dataset of \gls{csi} in five scenarios using off-the-shelf smartphones.
	\item We implement \name and evaluate it based on the collected data, demonstrating its accurate copresence detection, considering different frequency bands, heterogeneous devices, and attack scenarios. We also show the capability of \name to work reliably in real-time.
	\item We publicly release the collected dataset as well as the source code of our \gls{csi} data collection app, evaluation stack, and \name prototype. 
\end{itemize}


\section{Background and Related Work}
\label{sec:back-work}
\begin{figure}
    \centering
    \begin{minipage}[b]{0.39\textwidth}
    	\centering
    	\includegraphics[width=1\linewidth]{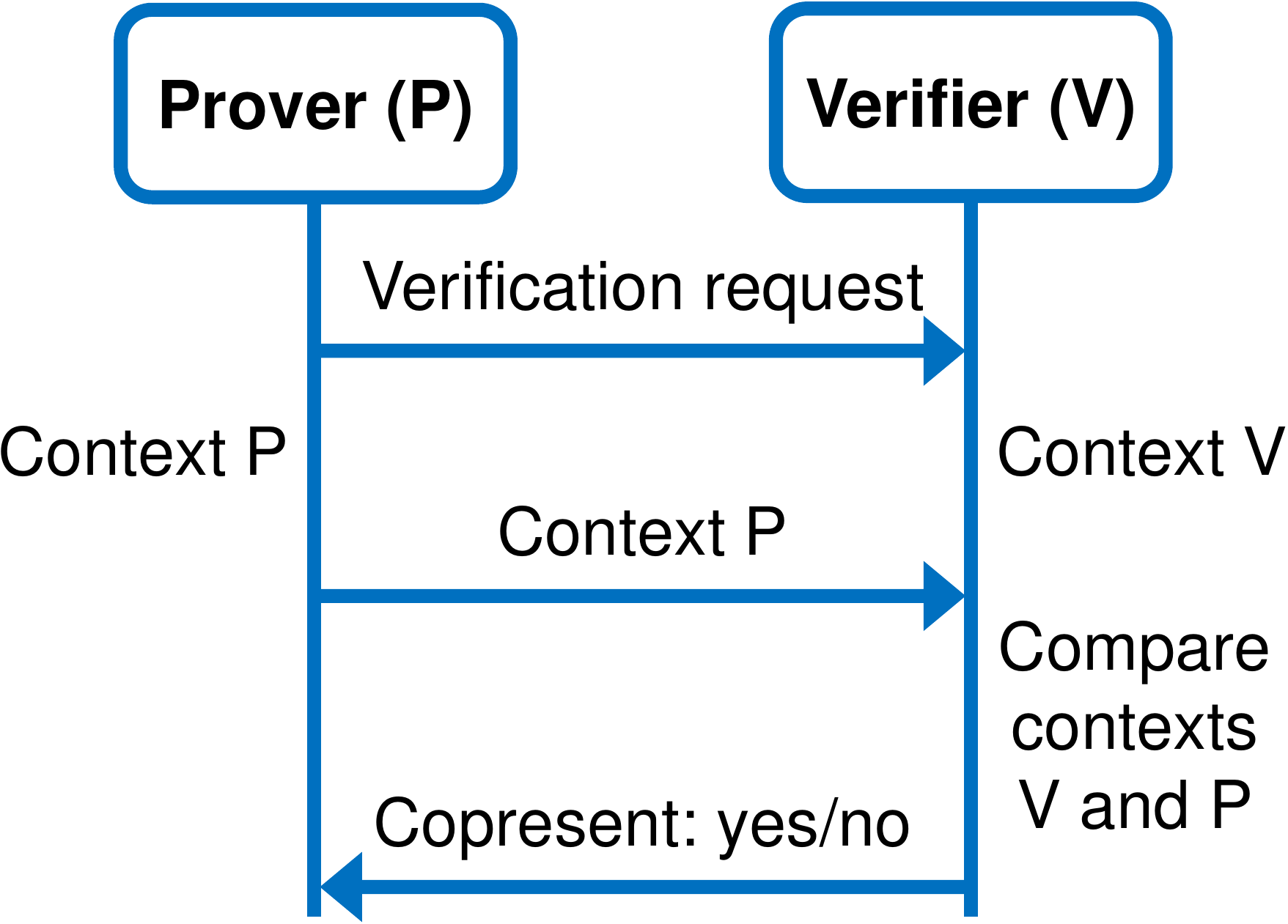}
    	\vspace{2.4mm}
    	\caption{Copresence detection flow between a prover and verifier.}
    	\label{fig:copres-flow}
    \end{minipage}
    \hfill
    \begin{minipage}[b]{0.56\textwidth}
    \small
    \centering
    \begin{adjustbox}{width=1\textwidth}
    \begin{forest}
        family tree too,
        [Wireless signals, l sep = 3mm, s sep = 2mm
                [Indoor localization, l=3mm, s sep = 2mm, l sep = 0mm
                            [Infrastructure-based, l=2mm, s sep = 0.3mm
							  	[\gls{rss}~\cite{feng2011received}, draw=none]
							  	[\gls{csi}~\cite{wang2016csi, wu2012csi}, draw=none]
							  	[\makecell{\gls{toa}~\cite{perez2019probabilistic} \\
							  	\gls{tdoa}~\cite{niitsoo2018convolutional} \\
							  	\gls{aoa}~\cite{xiong2013arraytrack}}, draw=none]
                            ]
                            [Ad-hoc, l=2mm, s sep = 0.3mm
                            	[\gls{rss}~\cite{abouzar2016rssi}, draw=none]
                            ]
                ]
              	[Copresence detection, l=3mm, s sep = 2mm, l sep = 0mm
                            [Infrastructure-based, l=2mm, s sep = 0.3mm
							  	[\gls{rss}~\cite{agata2015room, agata2016room}, draw=none]
                            ]
                            [Ad-hoc, l=2mm, s sep = 0.3mm
                            	[\textbf{\gls{csi} (\name)}, draw=none]
                            ]
               	]
            ]
    \end{forest}
    \end{adjustbox}
    \caption{Overview of indoor device localization and copresence detection schemes based on wireless signals.}
    \label{fig:loc-copres}
    \end{minipage}
\end{figure}

In this section, we explain how context-based copresence detection works and review existing schemes, demonstrating the advantages of \name. 
\\
\textbf{Background.} The goal of context-based copresence detection is to determine \textit{without the need of infrastructure} (e.g., \gls{ap}) whether two devices reside inside the same enclosed space (e.g., room or car). The two devices involved in the copresence detection task are called a \textit{prover} and \textit{verifier}, where the former tries to proof its physical proximity to the latter as follows (cf.~\autoref{fig:copres-flow}).
First, the prover sends a copresence verification request to the verifier over a wireless channel such as Bluetooth. Second, both devices capture their context using available sensors for a predefined timeframe (e.g., 10 seconds). Third, the prover transmits its context readings to the verifier over the wireless channel. This channel is secured by means of a shared key, thus the context readings are encrypted and authenticated, protecting them from adversaries.  
Such a shared key is assumed to be priorly established between the prover and verifier (e.g., via secure pairing~\cite{Fomichev:2017, putz2020acoustic}). 
Fourth, the verifier compares its context readings with the ones sent by the prover and decides if they are copresent.
To compare context readings, the verifier can either use similarity metrics (e.g., cross-correlation) and check them against the set thresholds~\cite{Karapanos:2015, Halevi:2012, Shepherd:2017} or compute features from context readings (e.g., median) and input them to a trained machine learning classifier~\cite{Truong:2014, Mare:2014, Shrestha:2014, Truong:2019, Shrestha:2019}. 
\\
\textbf{Related Work: Copresence Detection.}
To date, a number of context-based copresence detection schemes relying on various sensor modalities, including audio, radio signal strength, GPS as well as \gls{imu} (e.g., accelerometer, gyroscope) and environmental sensors (e.g., thermometer, barometer) have been proposed~\cite{Halevi:2012, Truong:2014, Mare:2014, Shrestha:2014, Karapanos:2015, Shepherd:2017, Truong:2019, Shrestha:2019}. 
Their details can be found in the survey by Conti and Lal~\cite{Conti:2019}. 
The existing schemes require devices to have common sensors such as microphones, limiting their applicability, whereas \name only needs a Wi-Fi chipset, which is ubiquitous in \gls{iot} devices. 
Recent works demonstrate that existing schemes have reduced copresence detection accuracy, hence lower security, in low-entropy context (e.g., empty room with few events occurring) and insufficiently separated environments (e.g., adjacent rooms)~\cite{Shrestha:2018, Fomichev:2019, Truong:2019}. 
Specifically, Fomichev et al.~\cite{Fomichev:2019} reproduce three state-of-the-art schemes, showing their vulnerability to the above threats using real-world data. 
Similarly, Shrestha et al.~\cite{Shrestha:2018} present successful context injection by a nearby adversary utilizing off-the-shelf home appliances. 
Truong et al.~\cite{Truong:2019} perform an efficient context manipulation attack, rendering their previous copresence detection scheme~\cite{Truong:2014} insecure.  
The aforementioned issues reveal an open challenge of accomplishing secure copresence detection without the need of shared sensors, which has motivated our scheme \name. 
In~\autoref{sec:eval}, we demonstrate that \name achieves accurate copresence detection in low-entropy context and insufficiently separated environments, mitigating the corresponding attacks and outperforming state-of-the-art copresence detection schemes.  
\\
\textbf{Related Work: Indoor Device Localization.} In contrast to copresence detection, indoor localization assumes the availability of infrastructure, where the goal is to find either fine- or coarse-grained location of devices\footnote{Fine-grained localization is considered to have submeter accuracy, while coarse-grained localization can have room-level precision (i.e., a device is located in room A or B)~\cite{yang2013rssi}.} given a reference point (e.g., wireless router); indoor device localization schemes utilizing various sensor modalities are surveyed by Xiao et al.~\cite{xiao2016survey}. 
However, we consider only those schemes that are based on wireless signals to enable a meaningful comparison with \name.

\autoref{fig:loc-copres} shows that the majority of proposed indoor localization schemes require infrastructure such as an \gls{ap} to operate, while existing ad hoc localization methods are mostly theoretical, still assuming the presence of trusted anchors~\cite{abouzar2016rssi}. 
Prior works rely on \gls{rss}~\cite{feng2011received}, \gls{csi}~\cite{wang2016csi, wu2012csi}, \gls{toa}~\cite{perez2019probabilistic}, \gls{tdoa}~\cite{niitsoo2018convolutional}, or \gls{aoa}~\cite{xiong2013arraytrack} to achieve device localization.
Specifically, for a device to be localized, it is required to (1) record \gls{rss}, \gls{csi}, \gls{toa}/\gls{tdoa}, or \gls{aoa} from multiple \glspl{ap} in its vicinity (i.e., 3--30) at 10--100 measurements per second, (2) often have the line of sight with these \glspl{ap}, and (3), in the case of \gls{toa}/\gls{tdoa} or \gls{aoa}, be equipped with multiple antennas.
Moreover, many of these schemes rely on either dedicated bulky hardware (e.g., software defined radios), requiring calibration, or they need heavy post-processing and filtering of the data to accurately estimate \gls{csi}, compensating for hardware imperfections, which are native to wireless chipsets in off-the-shelf devices. 
The above requirements (i.e., many measurements per second, multiple antennas, heavy data processing) are excessive for the majority of \gls{iot} devices, and existing localization schemes cannot be used in environments with few ambient \glspl{ap} such in vehicular or rural settings. 
On the contrary, \name can work on simple devices with a single antenna requiring only 2--3 \gls{csi} measurements per second, that do not need heavy post-processing, and without any infrastructure such as \glspl{ap} (cf.~\autoref{tab:loc-copres}).

Leveraging on the rich literature on device localization, infrastructure-based copresence
detection schemes utilizing \gls{rss} have also been proposed~\cite{agata2015room, agata2016room}. 
However, the applicability of these schemes is limited due to the strong assumption on the infrastructure availability. 
\\
\textbf{Practicality of RSS for Localization and Copresence Detection.}
Despite being a simple measurement and available on many devices, \gls{rss} is known to be inaccurate and unstable, especially in complex environments with many obstacles and human/object motion, hindering reliable indoor localization~\cite{yang2013rssi, wang2016csi, wu2012csi}. 
This finding is also confirmed by Fomichev et al.~\cite{Fomichev:2019} for the copresence detection scheme by Truong et al.~\cite{Truong:2014}, which uses the \gls{rss} measurements of ambient Wi-Fi \glspl{ap} and Bluetooth devices. 
Specifically, they demonstrate that \gls{rss} has low relevance for copresence detection\footnote{A possible explanation for this result is that \gls{rss} can be regarded as a single measurement obtained from the high-dimensional \gls{csi}, thus incurring in information loss.} not only in environments with scarce \glspl{ap} and Bluetooth devices (e.g., parking lot) but also in a busy office, where many \glspl{ap} and Bluetooth devices exist but the environment is complex.
Similarly, Zhu et al.~\cite{Zhu:2020} shows that using a single-antenna smartphone requires multiple rounds of \gls{rss} measurements and significant processing to ensure their quality in order to achieve satisfactory indoor localization accuracy. 
Furthermore, since \gls{rss} represents the power of the received signal, attacking any
localization or copresence detection scheme becomes trivial by simply increasing the power of adversarial devices. This threat is not considered by the existing schemes but is shown to be feasible by Zhu et al.~\cite{Zhu:2020} using off-the-shelf hardware. 
\name, on the other hand, is only limitedly affected by the increased power attack, as we demonstrate in~\autoref{subsec:adv}.
\\
\textbf{Related Work: Other CSI Applications.} 
Other works utilize \gls{csi} for more elaborate tasks than copresence detection like user identification/authentication~\cite{Shi:2017} or even imaging~\cite{huang2014feasibility}, which can be repurposed to detect copresence. 
However, these techniques impose even higher requirements in terms of \gls{csi} measurements per second (e.g., up to 1000), number of antennas, calibration, and complexity of the data processing than the indoor localization schemes described above. 
Thus, it is unlikely that such techniques can be practically used on \gls{iot} devices for copresence detection. 

In summary, \name stands in contrast to the above copresence detection, indoor localization, and other \gls{csi}-based schemes being the first ad hoc copresence detection scheme that utilizes \gls{csi}, imposing minimum requirements (e.g., few CSI measurements per second, single antenna) on devices executing \name and addressing advanced attacks such as an increased power attack.

\begin{table}
\small
\centering
	\caption{Comparison between operating requirements of indoor localization schemes and \name.}
	\label{tab:loc-copres}
  \begin{tabular}{l|ccccc}
  	\toprule
  	Approach & \makecell{Number of \\ \glspl{ap}} & \makecell{\gls{csi} measurements \\ per second} & \makecell{Number of \\ antennas} & \makecell{Line-of-sight \\ transmission} & \makecell{Processing \\ overhead} \\
  	\midrule
  	Localization & 3--30 & 10--100 & Multiple & Often yes & Medium--high \\
  	\name & 0 & 2--3 & One & No & Low \\
  	\bottomrule
  \end{tabular}
\end{table}

\section{System and Threat Models}
\label{sec:models}
In this section, we present our system model, describing the goal, requirements, and assumptions of \name as well as our threat model, detailing adversary's goal and capabilities. 
\\
\textbf{System Model.}
The main \textit{goal} of \name is for one device (\textit{prover}) to prove its copresence within a trusted boundary (e.g., inside a room) to another device (\textit{verifier}) using their context.
We design \name to fulfill the following \textit{requirements}: (1) be free of user interaction (\textit{usability}), (2) provide reliable copresence detection in low-entropy contexts and insufficiently separated environments (\textit{robustness}), and (3) work on  off-the-shelf devices such as smartphones equipped only with a Wi-Fi chipset (\textit{deployability}). 
To achieve the set goal while satisfying the requirements, we make the following \textit{assumptions}: (1) the prover
can send Wi-Fi frames to the verifier, which extracts \gls{csi} upon frame reception, (2) the prover and verifier share a secret key, allowing the verifier to ensure frame origin, mitigating replay attacks (e.g., by using a random nonce encrypted with a shared key). 
\\
\textbf{Threat Model.}
The \textit{goal} of the adversary is to convince the verifier that they are copresent, while not being located within the trusted boundary. Specifically, the adversary aims to either \textit{impersonate} a legitimate prover or launch a \textit{relay attack}, where a pair of colluding adversaries forward messages between the prover and verifier. 
We assume that the adversary can neither compromise the verifier or legitimate prover nor break the encryption between them. 
However, the adversary fully controls the wireless channel (i.e., can drop, modify, or replay frames) and uses the following capabilities to achieve their goal. 
In the first case (\textit{passive attack}), the adversary located outside the trusted boundary sends Wi-Fi frames to the verifier, triggering \gls{csi} extraction.
The adversary is equipped with similar off-the-shelf hardware (e.g., smartphones) to the prover's, thus they can stealthily deploy their devices right outside the trusted boundaries (e.g., in the adjacent office), increasing the attack surface~\cite{Zhu:2020}. 
In addition, the adversary can move their devices along the perimeter of the trusted boundary and stay there for a prolonged time, including periods of low-entropy context (e.g., adjacent offices at night).
In the second case (\textit{active attack}), the adversary has all the capabilities as in the passive attack and can additionally manipulate transmission of Wi-Fi frames such as sending frames of different types or increasing transmission power, aiming to match their \gls{csi} with that of legitimate devices inside the trusted boundaries. 

We consider adversaries copresent with legitimate devices (e.g., within the same room) to be outside the scope of this work.
Prior research shows that for audio such attacks can be mitigated, however, it limits copresence detection to distances below half a meter apart \cite{Truong:2019}. 
In the case of \gls{csi}, a pairing scheme exhibits a similar trade-off: pairing devices must be no further than a few centimeters apart if the adversary is copresent \cite{Xi:2016}.  


\section{System Design}
\label{sec:sysdesign}
In this section, we first explain the rationale for using \gls{csi} together with neural networks for copresence detection and then present the architecture of \name, describing its modules: \textit{data collection}, \textit{data processing}, and \textit{copresence decision-making}. 

\begin{figure}
	\begin{center}
		\includegraphics[width=0.995\linewidth]{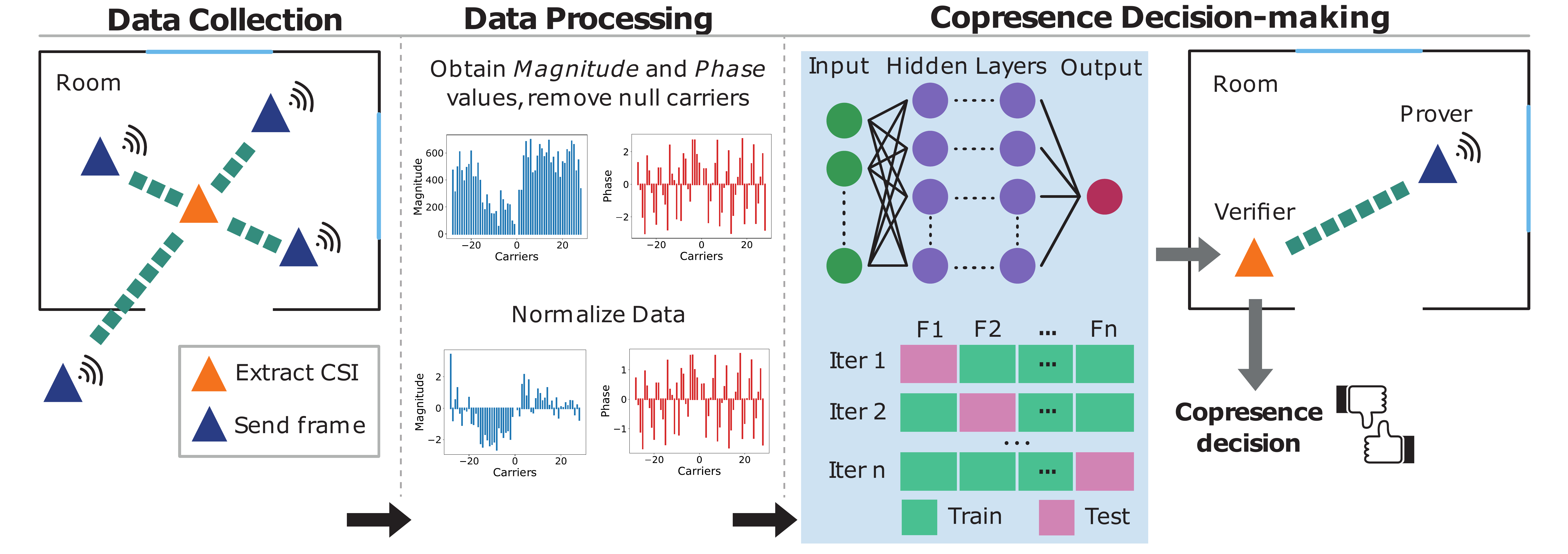}
	\end{center}
	\caption{System overview. \name uses a representative dataset of processed \gls{csi} data to train a cross-validated neural network model that is deployed on a verifier device to predict copresence in real-time.}
	\label{fig:sys-overview}
\end{figure}

\subsection{System Overview}
\label{subsec:sys-overview}
The main goal of \name is to allow the prover device to confirm its copresence to the verifier device. 
We consider the following use cases of \name: it determines copresence of devices located inside the same room or car.
Such use cases are typical in the \gls{iot}, for example, in a smart home, devices are often moved between different rooms (e.g., smart lamp). Once the relocated device is deployed, the room's smart hub can automatically provide access to the local subnetwork based on copresence. Similarly, in a connected car, passengers' smartphones can seamlessly share content with the infotainment system due to their copresence. 

\name achieves copresence detection in three steps (cf.~\autoref{fig:sys-overview}). First, the \textit{data collection} module allows the verifier to obtain representative \gls{csi} data collected by multiple devices at different spots inside a room or car.
This is feasible because the number of \gls{iot} devices already reaches roughly a dozen per household~\cite{Aviva:2020}, increasing to hundreds in few years~\cite{Han:2018}, thus each room in a smart home will be densely covered by distributed \gls{iot} devices; the same trend is observed for connected cars~\cite{Shuman:2019}. 
Second, the \textit{data processing} module converts the collected \gls{csi} data into magnitude and phase values of Wi-Fi subcarriers, removes irrelevant values, and performs data normalization. 
Third, the processed \gls{csi} data is input to the \textit{copresence decision-making} module to train and validate a neural network model that is deployed on the verifier. 
Afterwards, the prover wishing to confirm its copresence sends a number of Wi-Fi frames to the verifier, which extracts \gls{csi} data upon frame reception, processes it as described above, and inputs the processed \gls{csi} to the trained  neural network model that outputs a copresence decision. 

\subsection{Rationale for Using \gls{csi} and Neural Networks for Copresence Detection}
\label{subsec:csi-rationale}
In the following, we (1) justify the suitability of \gls{csi} for copresence detection, (2) rationalize why neural networks can best leverage location-sensitive properties of \gls{csi} to detect copresence, and (3) motivate the choice of Wi-Fi standards that we use for \name experimentation. 
\\
\textbf{\gls{csi} as a Copresence Feature.}
We demonstrate that \gls{csi} is useful for copresence detection because of two reasons. First, it provides a discretized frequency-domain representation of the \gls{cir}, which captures a wireless fingerprint of an environment (e.g., a room) in terms of path-loss, fading, reflections, and scattering of the wireless channel~\cite{Ma:2019}.
Second, \gls{csi} is the default information generated when two Wi-Fi devices communicate, and it becomes increasingly available in off-the-shelf devices such as routers, laptops, and smartphones (cf.~\autoref{sec:disc}). 
In the following, we provide an expression relating \gls{csi} and \gls{cir} via a linear transform, demonstrating that the former indirectly measures the latter, and thus captures a wireless fingerprint of an environment.

\def\cont{(-0.105, 0)
  -- (28.3, 0)
  -- (28.3, 26.3) 
  -- +(-4.7, 0)
  -- +(-4.7, 4.2)
  -- (3.6, 30.5)
  -- +(0, -4.5)
  -- +(-3.6, -4.5)
  -- (0, 0)}%

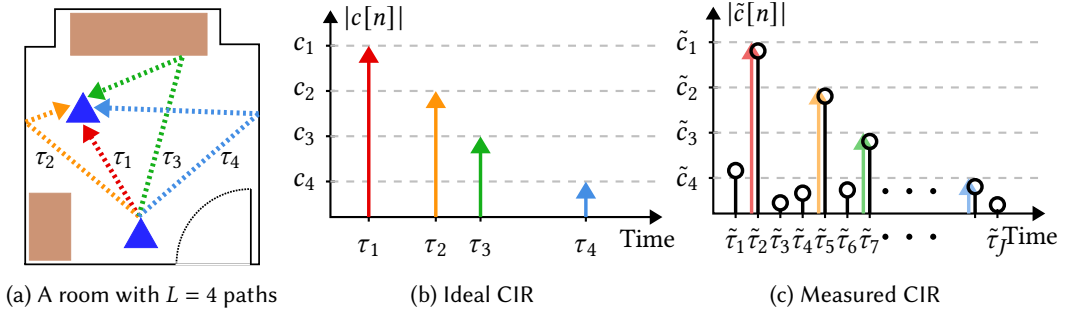
\begin{figure}
\begin{subfigure}[b]{0.26\textwidth}
	\centering
	\scalebox{.11}{
	\begin{tikzpicture}[
	  furniture/.style = {
	    ultra thick,
	    fill = AntiqueBrass
	  }]
	
	  \draw[line width=6pt] \cont;
	
	  \draw[line width=6pt, loosely dotted] (18.3, 0) arc (180:90:9);
	  \draw[line width=6pt] (18.3 + 9, 0) -- +(0, 9);
	  \draw[line width=6pt, white] (18.192, 0) -- +(9, 0);
		
	  \begin{scope}[xshift = 0.5cm, yshift = 8.5cm, rotate = -90]
	    \draw[AntiqueBrass, furniture] (0, 0) rectangle (8, 5);
	  \end{scope}
	
	  \begin{scope}[xshift = 22cm, yshift = 25cm, rotate = 90]
	    \draw[AntiqueBrass, furniture] (0, 0) rectangle (5, 16.5);
	  \end{scope}
	
	  \begin{scope}[
	    chair/.style = {
	      ultra thick,
	      fill = red,
	      rounded corners = 3em
	    }]
	    
	    \draw[blue!85!white, fill=blue!85!white] (5, 17)--(9, 17)--(7, 20.5)--cycle;   
	    \draw[blue!85!white, fill=blue!85!white] (12, 2)--(16, 2)--(14, 5.5)--cycle;
	    
	    \draw[line width=6mm, >=Triangle, ->, dash pattern=on 10pt off 12pt, myred] (13.75, 5.75) to (7, 16.75);
	    
	    \draw[line width=6mm, dash pattern=on 10pt off 12pt, mytangerine] (13.75, 5.75) to (0.05, 17);
	    \draw[line width=6mm, >=Triangle, ->, dash pattern=on 10pt off 12pt, mytangerine] (0.05, 17) to (5.5, 19);
	    
	    \draw[line width=6mm, dash pattern=on 10pt off 12pt, mygreen] (13.75, 5.75) to (19, 24.75);
	    \draw[line width=6mm, >=Triangle, ->, dash pattern=on 10pt off 12pt, mygreen] (19, 24.75) to (7.5, 20);
	    
	  	\draw[line width=6mm, dash pattern=on 10pt off 12pt, mysky] (13.75, 5.75) to (28.25, 18);
	    \draw[line width=6mm, >=Triangle, ->, dash pattern=on 10pt off 12pt, mysky] (28.25, 18) to (8.25, 18.75);
	    
	    \node[font=\fontsize{95}{115}] at (11.8, 12.5) {$ \tau_\text{1} $}; 
	    
	    \node[font=\fontsize{95}{115}] at (2.35, 12.5) {$ \tau_\text{2} $};
	    
	    \node[font=\fontsize{95}{115}] at (17.75, 12.5) {$ \tau_\text{3} $};
	    
	    \node[font=\fontsize{95}{115}] at (24.75, 12.5) {$ \tau_\text{4} $};
	    	    
	  \end{scope}
	
	\end{tikzpicture}
	}
   	
   	\caption{A room with $L = 4$ paths}
   	\label{sf:cir_scenario}
\end{subfigure}
\begin{subfigure}[b]{0.36\textwidth}
	\centering
	\pgfplotsset{grid style={dashed}}
  	\begin{tikzpicture}[scale=.85]
  	
  	\draw [>=Triangle, <->, thick] (0, 3.2) node (yaxis) {} |- (5.2, 0) node (xaxis) {};
  	
  	\draw [thick] (2pt, 0.55) -- (-2pt, 0.55) node[anchor=east] {\large $ c_4 $}; 
  	\draw [thick] (2pt, 1.25) -- (-2pt, 1.25) node[anchor=east] {\large $ c_3 $}; 
  	\draw [thick] (2pt, 1.95) -- (-2pt, 1.95) node[anchor=east] {\large $ c_2 $};
  	\draw [thick] (2pt, 2.65) -- (-2pt, 2.65) node[anchor=east] {\large $ c_1 $};
  	
	\draw [thick] (0.6, 2pt) -- (0.6, -2pt) node[anchor=north] {\large $ {\color{white}\tilde{\color{black}{\tau}}}_1 $};
	\draw [thick] (1.65, 2pt) -- (1.65, -2pt) node[anchor=north] {\large $ {\color{white}\tilde{\color{black}{\tau}}}_2 $};
	\draw [thick] (2.35, 2pt) -- (2.35, -2pt) node[anchor=north] {\large $ {\color{white}\tilde{\color{black}{\tau}}}_3 $};
	\draw [thick] (4.0, 2pt) -- (4.0, -2pt) node[anchor=north] {\large $ {\color{white}\tilde{\color{black}{\tau}}}_4 $};
  	
  	\node[xshift=-0.2cm, below=0.04cm] at (xaxis) {Time};
  	  	\node[yshift=-0.1cm, right=0.05cm] at (yaxis) {$ \left| c [n] \right| $};
  	
  	\draw[thick, dashed, gray!50] (0, 0.55) to (5.2, 0.55);
  	\draw[thick, dashed, gray!50] (0, 1.25) to (5.2, 1.25);
  	\draw[thick, dashed, gray!50] (0, 1.95) to (5.2, 1.95);
  	\draw[thick, dashed, gray!50] (0, 2.65) to (5.2, 2.65);
  	
  	\draw[line width=1.3pt, >=Triangle, ->, myred] (0.6, 0) to (0.6, 2.65);
  	\draw[line width=1.3pt, >=Triangle, ->, mytangerine] (1.65, 0) to (1.65, 1.95);
  	\draw[line width=1.3pt, >=Triangle, ->, mygreen] (2.35, 0) to (2.35, 1.25);
  	\draw[line width=1.3pt, >=Triangle, ->, mysky] (4.0, 0) to (4.0, 0.55);
  	
  	\end{tikzpicture} 
   	\caption{Ideal \gls{cir}}
    \label{sf:cir_ideal}
\end{subfigure}
\begin{subfigure}[b]{0.36\textwidth}
	\centering
	\pgfplotsset{grid style={dashed}}
  	\begin{tikzpicture}[scale=.85]

  	\draw [>=Triangle, <->, thick] (0, 3.2) node (yaxis) {} |- (5.2, 0) node (xaxis) {};
  	
  	\draw [thick] (2pt, 0.55) -- (-2pt, 0.55) node[anchor=east] {\large $ \tilde{c}_4 $}; 
  	\draw [thick] (2pt, 1.25) -- (-2pt, 1.25) node[anchor=east] {\large $ \tilde{c}_3 $}; 
  	\draw [thick] (2pt, 1.95) -- (-2pt, 1.95) node[anchor=east] {\large $ \tilde{c}_2 $};
  	\draw [thick] (2pt, 2.65) -- (-2pt, 2.65) node[anchor=east] {\large $ \tilde{c}_1 $};
  	
 	\draw [thick] (0.35, 2pt) -- (0.35, -2pt) node[anchor=north] {\large $ \tilde{\tau}_1 $}; 
 	\draw [thick] (0.7, 2pt) -- (0.7, -2pt) node[anchor=north] {\large $ \tilde{\tau}_2 $}; 
	\draw [thick] (1.05, 2pt) -- (1.05, -2pt) node[anchor=north] {\large $ \tilde{\tau}_3 $}; 
	\draw [thick] (1.4, 2pt) -- (1.4, -2pt) node[anchor=north] {\large $ \tilde{\tau}_4 $};
	\draw [thick] (1.75, 2pt) -- (1.75, -2pt) node[anchor=north] {\large $ \tilde{\tau}_5 $};
	\draw [thick] (2.1, 2pt) -- (2.1, -2pt) node[anchor=north] {\large $ \tilde{\tau}_6 $};
	\draw [thick] (2.45, 2pt) -- (2.45, -2pt) node[anchor=north] {\large $ \tilde{\tau}_7 $};
	
	\node[xshift=-1.75cm, below=-0.45cm] at (xaxis) {\Huge $\cdots$};
	\node[xshift=-1.75cm, below=0.15cm] at (xaxis) {\Huge $\cdots$};
	
	\draw [thick] (4.1, 2pt) -- (4.1, -2pt) node[anchor=north] {};
	\draw [thick] (4.45, 2pt) -- (4.45, -2pt) node[anchor=north] {\large $ \tilde{\tau}_J $};
  	
  	\node[xshift=-0.2cm, below=0.04cm] at (xaxis) {Time};
  	\node[yshift=-0.1cm, right=0.05cm] at (yaxis) {$ \left| \tilde{c} [n] \right| $};
  	
  	\draw[thick, dashed, gray!50] (0, 0.55) to (5.2, 0.55);
  	\draw[thick, dashed, gray!50] (0, 1.25) to (5.2, 1.25);
  	\draw[thick, dashed, gray!50] (0, 1.95) to (5.2, 1.95);
  	\draw[thick, dashed, gray!50] (0, 2.65) to (5.2, 2.65);
  	
  	\draw [line width=1.3pt, -{Circle[open]}](0.35, 0) to (0.35, 0.8);
	\draw[line width=1.3pt, >=Triangle, ->, myred!60] (0.6, 0) to (0.6, 2.65);
	\draw[line width=1.3pt, -{Circle[open]}] (0.7, 0) to (0.7, 2.65);
	\draw[line width=1.3pt, -{Circle[open]}] (1.05, 0) to (1.05, 0.3);
	\draw[line width=1.3pt, -{Circle[open]}] (1.4, 0) to (1.4, 0.45);
	\draw[line width=1.3pt, >=Triangle, ->, mytangerine!60] (1.65, 0) to (1.65, 1.95);
	\draw[line width=1.3pt, -{Circle[open]}] (1.75, 0) to (1.75, 1.95);
	\draw[line width=1.3pt, -{Circle[open]}] (2.1, 0) to (2.1, 0.5);
	\draw[line width=1.3pt, >=Triangle, ->, mygreen!60] (2.35, 0) to (2.35, 1.25);
	\draw[line width=1.3pt, -{Circle[open]}] (2.45, 0) to (2.45, 1.25);
  	\draw[line width=1.3pt, >=Triangle, ->, mysky!60] (4.0, 0) to (4.0, 0.55);
  	\draw[line width=1.3pt, -{Circle[open]}] (4.1, 0) to (4.1, 0.55);
  	\draw[line width=1.3pt, -{Circle[open]}] (4.45, 0) to (4.45, 0.27);
  	
  	\end{tikzpicture} 
   	\caption{Measured \gls{cir}}
    \label{sf:cir_measured}
\end{subfigure}
\caption{Channel impulse response (\gls{cir}) of a room. \autoref{sf:cir_scenario} shows propagation of the transmitted signal between two devices via $L = 4$ paths. \autoref{sf:cir_ideal} depicts the ideal continuous-time \gls{cir} of the scenario in~\ref{sf:cir_scenario}, while~\autoref{sf:cir_measured} shows its discrete-time version, occurring due to limitations in sampling and bandwidth.}
\label{fig:cir}
\end{figure}

A signal propagating through an environment such as a room experiences changes that are distinctive to physical characteristics of the surroundings. 
Such changes---observed as magnitude and phase variations---characterize the communication channel, capturing geometry of the environment, distribution of objects within, and nature of the materials. 
Mathematically, the communication channel between a transmitter and receiver is represented by the \gls{cir}, which \textit{models the overall effect of reflectors, absorbers, path-loss, and complexity of the environment between them}.
\autoref{sf:cir_scenario} shows an example of two devices communicating inside the same room that we use for our explanation. 
We denote the continuous-time \gls{cir} of an $ L $-path baseband wireless communication channel as:
\begin{equation} \label{eq:ideal-cir}
	c(t) = \sum^L_{i = 1} c_i \delta(t - \tau_i).
\end{equation}

In~\autoref{eq:ideal-cir}, $ \delta(t - \tau_i) $ is the Dirac delta function representing a delayed multi-path replica of the transmitted signal arriving at time $ \tau_i $ with power $ \left| c_i \right|^2 $. 
In particular, $ c_i = a_i e^{j \theta_i} $, where $ a_i $ and $ \theta_i $ denote the amplitude and phase of the $ i $-th replica, as shown in~\autoref{sf:cir_scenario} and~\autoref{sf:cir_ideal}. 
We note that $ c(t) $ fully describes the communication channel between the transmitter and receiver. 
Nevertheless, there exist technical challenges in the wireless communication chain that hinder its accurate acquisition. Specifically, limitations in the sampling frequency and bandwidth incur in information loss, thereby preventing accurate knowledge of $ c(t) $. 
As a result, only a surrogate version of the \gls{cir} can be obtained, which is expressed as:
\begin{equation} \label{eq:discrete-cir}
 c \left[ n \right] = \sum^L_{i = 1} c_i \frac{\sin \left( \pi \left( n \cdot \Delta \tau - \tau_i \right)  \right) }{\pi \left( n \cdot \Delta \tau - \tau_i \right) }.
\end{equation}

Note that the discrete-time \gls{cir} in~\autoref{eq:discrete-cir} is a sampled version of the ideal continuous-time \gls{cir} in~\autoref{eq:ideal-cir}. 
In particular, $ \Delta \tau = \frac{1}{B} $ represents the time resolution (i.e., spacing in seconds between samples), which is inversely proportional to the channel bandwidth $ B $. 
If all $ \tau_i $ are multiples of $ \Delta \tau $ then $ c \left[ n \right] $ and $ c(t) $ become equivalent. 
Otherwise, if some $ \tau_i $ is not a multiple of $ \Delta \tau $, the energy of that element spreads across all the elements $ c \left[ n \right] $ (i.e., energy leakage) owing to oscillations of the sampling function $ \frac{\sin \left( \pi \left( n \cdot \Delta \tau - \tau_i \right)  \right) }{\pi \left( n \cdot \Delta \tau - \tau_i \right) } $, which produce artificial small-valued samples, as depicted in~\autoref{sf:cir_measured}. \textit{In general, due to sampling and bandwidth limitations, energy leakage inevitably occurs among the samples of $ c \left[ n \right] $.} Thus, a more realistic representation of the measured CIR is given by:
\begin{equation} \label{eq:general-cir}
	\tilde{c} \left[ n \right] = \sum^J_{i = 1} \tilde{c}_i \delta(n \cdot \Delta \tau - \tilde{\tau}_i).
\end{equation}

Here, all $ \tilde{\tau}_i $ are multiples of $ \Delta \tau $, and $ \tilde{c} \left[ n \right] $ represents a discrete-time distorted version of $ c(t) $, which may not only include perturbations due to sampling and bandwidth limitations but also due to amplitude quantization. 
We see that $ \tilde{c} \left[ n \right] $ in~\autoref{sf:cir_measured} has $ L = 4 $ prominent paths similarly to $ c(t) $ except for the additional spurious small-valued samples.
Although $ c(t) $ cannot be completely captured, due to the reasons stated above, there is still valuable information in $ \tilde{c} \left[ n \right] $, which approximately describes the propagation environment. 
Thus, using \gls{cir} (or more precisely $ \tilde{c} \left[ n \right] $) for copresence detection is a sound strategy. 
However, in the case of Wi-Fi that uses OFDM, the \gls{cir} is not readily available. 
Instead, every Wi-Fi device measures \gls{csi}, as it is required for channel estimation and equalization. 
Fortunately, in OFDM systems, \gls{csi} and \gls{cir} are related by a bijective mapping through the discrete Fourier transform (DFT) matrix (cf.~\autoref{eq:csi-cir}).
It means that the matrix $ \mathbf{F} $ is invertible, ensuring one-to-one correspondence between \gls{cir} and \gls{csi}. 
Thus, for a given \gls{csi} measurement of a Wi-Fi frame with $ K $ subcarriers, the resulting \gls{cir} is unique, and vice versa.
\begin{equation} \label{eq:csi-cir}
		\underbrace{\begin{pmatrix}
		H_1 \\
		H_2 \\
		\vdots \\
		H_K
		\end{pmatrix}}_\mathrm{\gls{csi}}
		=
		\underbrace{\begin{pmatrix}
			1		& 1				& \cdots					& 1 \\
			1		& \exp\left( -j \frac{2 \pi}{K} \right)		& \cdots		& \exp\left( -j \frac{2 \pi (K-1)}{K} \right) \\
			\vdots  & \vdots	    & \ddots     	& \vdots  \\
			1		& \exp\left( -j \frac{2 \pi (K-1)}{K} \right) & \cdots		& \exp\left( -j \frac{2 \pi (K-1)(K-1)}{K} \right) \\
		\end{pmatrix}}_{\mathbf{F}: \text{~discrete Fourier transform matrix}}
		\underbrace{\begin{pmatrix}
			\tilde{c}_1 \\
			\tilde{c}_2 \\
			\vdots \\
			\tilde{c}_K
		\end{pmatrix}}_\mathrm{\gls{cir}}
\end{equation}

The \gls{csi} at the  $ k $-th subcarrier is defined as $ H_k = \sum^{K}_{n=1} \tilde{c}_n \exp\left( -j \frac{2 \pi (k-1) (n-1)}{K} \right) $, showing that the CSI at every subcarrier is a linear combination of all \gls{cir} elements. 
While computing the \gls{cir} is possible, it requires additional processing (i.e., inverse discrete Fourier transform), which will be computationally expensive on commodity devices. 
In contrast, \gls{csi} is inherently computed by every Wi-Fi device, capturing the same characteristics of the environment as the \gls{cir}. 
\\
\textbf{Neural Networks: Leveraging \gls{csi} for Copresence Detection.}
Despite capturing distinct characteristics of an environment (e.g., a room), \gls{csi} is sensitive to changes within such as motion and relocation of obstacles~\cite{Ma:2019, Shi:2017, Zhu:2020}. 
To enable copresence detection, we need to obtain unique features of the environment embedded in \gls{csi}, ensuring their robustness to insignificant changes. 
Hence, we require a technique that can accurately approximate data, capturing its distinct features but tolerating certain noise. 
To the best of our knowledge, neural networks best fulfill this purpose, in addition to their other advantages (cf.~\autoref{subsec:sd-ml}). 

We find that two points need to be considered to leverage \gls{csi} by neural networks for copresence detection.
First, \gls{csi} captures characteristics of the environment from the viewpoint of a transmitter-receiver pair, thus we need to provide a neural network with \gls{csi} observations from different spots inside the environment to obtain its general picture. 
Collecting \gls{csi} in multiple spots is feasible due to the growing number of \gls{iot} devices equipped with Wi-Fi (cf.~\autoref{subsec:sys-overview}). 
Second, for copresence detection, manually computed features from \gls{csi} frequently used by prior work (e.g., mean, power)~\cite{Ma:2019} perform worse in the neural network than raw magnitude and phase values. 
This happens because prior works engineer features capturing subtle \gls{csi} variations to detect a specific location, human, or activity. 
Such features inevitably reduce the amount of useful information in \gls{csi}, hindering the generalization capability of neural networks. 
The feature computation requires extra processing and more \gls{csi} data, and it prevents the representation and transfer learning provided by neural networks (cf.~\autoref{subsec:sd-ml}). 
We design \name guided by the above two points, demonstrating the capability of our neural network to utilize the rich environment information embedded in \gls{csi} to automatically learn robust copresence features (cf.~\autoref{sub:robust}). 
\\
\textbf{Wi-Fi Standards Used in \name.}
To demonstrate the practicality of \name, we utilize IEEE 802.11n (at 2.4 GHz) and IEEE 802.11ac (at 5 GHz) Wi-Fi standards. 
We choose them because of their favorable characteristics (described next) and ubiquity in various devices, ranging from simple sensors to powerful routers. 
First, the lower carrier frequency of 2.4 GHz in IEEE 802.11n enables communication between distant devices due to its robustness to path-loss and blockage.
In contrast, the higher carrier frequency of 5 GHz in IEEE 802.11ac makes it more vulnerable to such phenomena but allows capturing subtle details of the environment due to its shorter wavelength. 
Second, the narrower channel bandwidth of IEEE 802.11n (i.e., 20 MHz and 40 MHz) reduces the circuity requirements (e.g., 64-point and 128-point FFT chipsets), making it suitable for low-power \gls{iot} devices.
In contrast, the broader channel bandwidth of IEEE 802.11ac (i.e., up to 160 MHz) requires more expensive chipsets (e.g., 512-point FFT chipsets) but provides higher data rates for end-user devices such as laptops.  

\subsection{Data Collection}
To obtain a realistic \gls{csi} dataset in an environment such as a room, we collect \gls{csi} from several copresent devices located inside it at different spots (e.g., on a desk, window sill), varying in terms of nearby obstacles and height above the floor.
Similarly, we deploy a number of non-copresent devices outside the environment such as in an adjacent room, collecting \gls{csi} data from them as well. 
Thus, we obtain positive and negative copresence samples in comparable environments that are nearby. 
This allows a neural network to learn features that are common for copresent and distinct for non-copresent devices, considering the proximate environments.
Both copresent and non-copresent devices are fixed, however, we introduce dynamics to their environments by having people within, who frequently move, change position of obstacles (e.g., relocate a chair, open a door), and use other Wi-Fi devices such as laptops and \glspl{ap}. 
The data collection is finished when a copresent device in each spot have recorded \gls{csi} data from other copresent devices and all non-copresent devices for several minutes. 

\subsection{Data Processing}
We convert the raw \gls{csi} of a Wi-Fi frame to magnitude and phase shift values. 
Specifically, the magnitude $ M_k $ and phase $ \phi_k $ of the CSI at the $ k $ subcarrier (denoted by $ H_k $), are given by modulus and argument, as shown in \autoref{eq:mag-phase}.

\begin{equation} \label{eq:mag-phase}
M_k = \sqrt{ \mathfrak{Re} \left\lbrace H_k \right\rbrace^2 + \mathfrak{Im} \left\lbrace H_k \right\rbrace ^2},\quad\ \phi_k = \mathrm{atan} \left( \frac{\mathfrak{Im} \left\lbrace H_k \right\rbrace}{\mathfrak{Re} \left\lbrace H_k \right\rbrace} \right).
\end{equation}
Here, $ \mathfrak{Re} \left\lbrace H_k \right\rbrace $ and $ \mathfrak{Im} \left\lbrace H_k \right\rbrace $ are the real and imaginary parts of $ H_k $. The number of subcarriers depends on  the Wi-Fi channel bandwidth. For example, the 20 MHz channel in 802.11n consists of 64 subcarriers, resulting in 128 magnitude and phase values. 
In \name, we consider all subcarriers of a Wi-Fi channel in order to obtain a finer-grained wireless fingerprint of an environment (e.g., a room) captured by \gls{csi}.  
For example, similar to Shi et al.~\cite{Shi:2017}, we find that some subcarriers are more susceptible to noise than others with no discernible pattern observed. 
Such noise susceptibility might be distinctive to the environment, indicating a specific interference behavior. 
However, not all subcarriers provide meaningful \gls{csi}, namely \textit{null subcarriers}, which do not carry any information~\cite{Gawlowicz:2019}. 
Thus, we remove magnitude and phase values of such subcarriers from our data. 
We normalize the computed magnitude and phase values to have the same range, making them suitable to train a machine learning classifier on.
The structure of the processed \gls{csi} dataset is arranged in a matrix $ \mathbf{X} \in \mathbb{R}^{N \times D} $, as shown in \autoref{fig:csi-struct}, where $ D = 2 K' $ is the dimension of the feature vector for every measurement $ n = \left\lbrace 1, \cdots, N \right\rbrace $, $ K' $ is the number of useful subcarriers, while $ M_{n,k} $ and $ \phi_{n,k} $ denote the magnitude and phase of the $ k $-th subcarrier in the $ n $-th measurement.

\begin{figure} 
	\begin{equation*}
	\mathbf{X} = 
	\left (
	\begin{array}{cccc|cccc}
	M_{1,1} & M_{1,2} & \cdots & M_{1,K'} & \phi_{1,1} & \phi_{1,2} & \cdots & \phi_{1,K'} \\
	M_{2,1} & M_{2,2} & \cdots & M_{2,K'} & \phi_{2,1} & \phi_{2,2} & \cdots & \phi_{2,K'} \\
	\vdots & \vdots & \ddots & \vdots & \vdots & \vdots & \ddots & \vdots \\
	\undermat{\text{\gls{csi} magnitude}}{M_{N,1} & M_{N,2} & \cdots & M_{N,K'}} & \undermat{\text{\gls{csi} phase}}{\phi_{N,1} & \phi_{N,2} & \cdots & \phi_{N,K'}}
	\end{array}
	\right )
		\footnotesize
		\begin{array}{c}
		\text{\gls{csi} measurement} ~ 1 \\
		\text{\gls{csi} measurement} ~ 2 \\
		\vdots \\
		\text{\gls{csi} measurement} ~ N \\
		\end{array}
	\end{equation*}
	\vspace{3mm}	
\caption{Structure of processed \gls{csi} data. $ K' $ is the number of subcarriers of a Wi-Fi channel excluding null carriers, $N$ is the number of \gls{csi} measurements; $M_{n,k}$ and $\phi_{n,k}$ are normalized to be in the same range.}
\label{fig:csi-struct}
\end{figure}

\subsection{Copresence Decision-making}
\label{subsec:sd-ml} 
To capture a wireless fingerprint commonly observed by copresent devices inside the same environment (e.g., a room), we input the processed \gls{csi} data to a machine learning classifier.
Differently from existing copresence detection schemes utilizing machine learning~\cite{Halevi:2012, Truong:2014, Mare:2014, Shrestha:2014, Shepherd:2017, Truong:2019, Shrestha:2019}, we choose to use neural networks for the following reasons.  
First, neural networks under mild assumptions are universal function approximators, thus they have the potential for representing the classification function we are interested in learning~\cite{hornik1989multilayer,lu2017expressive,sonoda2017neural}.
Second, neural networks allow representation learning~\cite{Goodfellow-et-al-2016}, which replaces the manual feature engineering process, simplifying the modeling assumptions, saving time, and increasing accuracy.
The prior work on wireless signal classification demonstrates that neural networks are capable of learning the right representation for this domain, producing high predictive accuracy results~\cite{shi2019deep,o2018over,matuszewski2017neural,hauser2017signal}.
Third, the representation learning of neural networks enables transfer learning~\cite{tan2018survey}, where we can reuse the representation learned in one problem and embed it into the solution of another.
This has become a standard practice in the deep learning community, where big networks are trained on massive amounts of data. Such pretrained networks are publicly shared with other researchers and practitioners who adapt them to new tasks without training from scratch, significantly lowering the computational costs. 
In~\autoref{sub:general}, we leverage the capability for representation and transfer learning to demonstrate that our neural network can be adapted to new environments while reducing computational costs, making model training in \name feasible on battery-powered devices. 
Fourth, numerous deep learning frameworks, support for different devices, and constant improvements in neural networks facilitate the deployment of \name. 


\section{Implementation}
\label{sec:impl}
In this section, we provide the implementation details of our \gls{csi} collector, copresence decision-making module, and \name prototype. 

\subsection{\gls{csi} Collector}
\label{subsec:data-app}
For data collection, we develop an Android app, utilizing the \textit{Nexmon framework} \cite{Nexmon:2017} to extract \gls{csi}. 
Nexmon allows modifying the firmware of Broadcom Wi-Fi chipsets of Nexus 5 and Nexus 6P smartphones. 
We customize the original \gls{csi}-extractor~\cite{Schulz:2018} to our needs, allowing us to conduct experiments with both Nexus 5 and Nexus 6P devices. 

Our app works in two modes: the \textit{prover} and \textit{verifier}. 
The former broadcasts Wi-Fi frames at a predefined rate, while the latter listens for these frames, extracting \gls{csi} upon frame reception and storing it on a smartphone.
The app allows \gls{csi} collection in both 2.4 GHz and 5 GHz Wi-Fi bands. 
As input, the number and bandwidth of a Wi-Fi channel on which the prover and verifier communicate needs to be provided. 
We experiment with a number of Wi-Fi channels in both frequency bands using two criteria: (1) the stability of \gls{csi} collection and (2) maximum transmission range.
We find that many channels in the 2.4 GHz band vary significantly in terms of \gls{csi} collection stability, while the 5 GHz channels show a wide spread of transmission ranges. 
Our results obtained in various environments demonstrate that the channel 1 (20 MHz bandwidth, 2.4 GHz band) and channel 157 (80 MHz bandwidth, 5 GHz band) best satisfy the above criteria, thus we use them to collect \gls{csi} data in our experiments. 
With these channels, we obtain 128 and 510 magnitude and phase values from a single \gls{csi} measurement for 2.4 GHz and 5 GHz bands, respectively, which corresponds to using all subscarries of a Wi-Fi channel, including null ones. 
Furthermore, we implement different types of frames (i.e., \gls{qos} and beacon) used by the prover and verifier as well as varying transmission power of the prover, allowing us to evaluate the robustness of \name and its resilience to  attacks.
Using frames of around 100 bytes in size, we find that Nexmon allows extracting \gls{csi} at the rate of up to 20 frames per second. 
In our experiments, we use several provers and a single verifier (cf.~\autoref{subsec:exp-setup}), thus we set the transmission rate of a prover to three frames per second, ensuring reliable \gls{csi} data collection on the verifier.  

In addition, we port the functionality of the above app in the verifier mode onto the Raspberry Pi 3 Model B+ using the recent Nexmon \gls{csi} patch\footnote{\url{https://github.com/seemoo-lab/nexmon_csi}}. 
This allows us to further explore the capability of \name to perform on heterogeneous devices. 

\subsection{Copresence Decision-making}
\label{subsec:ml-pipeline}
We treat the problem of copresence detection as a binary classification task. 
Thus, we use the collected \gls{csi} data to train a machine learning classifier, predicting whether two devices are copresent (i.e., in the same room or car). 

To collect \gls{csi}, we change in turns the position of the verifier (e.g., on a window sill, on a desk), resulting in a number of \gls{csi} datasets from different verifier-provers combinations (cf.~\autoref{subsec:exp-setup}). 
To estimate the performance of our classification approach, we use a 5-fold \gls{cv}. This is a reasonable compromise between the computational costs of training the model and a good estimation of the predictive performance. 
Due to a larger number of non-copresent devices in our experiments, our dataset has an imbalanced distribution of the classes. 
Hence, we employ a stratified \gls{cv} and classification metrics that are not affected by the imbalanced data (cf.~\autoref{subsec:exp-setup}). 
We construct the folds by (1) loading the \gls{csi} data from all verifier-provers combinations, (2) shuffling it with a set seed (i.e., 123), and (3) evenly sampling the data into each fold.
Thus, we obtain five folds, each containing the same amount of data and ratio of positive and negative samples as the original \gls{csi} datasets from all verifier-provers combinations.
Prior to training, we delete the irrelevant \textit{null subcarriers} from the \gls{csi} data, resulting in 112 and 484 features (i.e., magnitude and phase values) per \gls{csi} measurement for 2.4 GHz and 5 GHz bands, respectively (cf.~\autoref{fig:csi-struct}). 
We normalize features in training and test sets using variance scaling~\cite{Zheng:2018}, ensuring that they have the same range, and thus are equally weighted by a classifier. The normalized features are obtained as:
\begin{equation}
x_{i, \mathrm{norm}} = \dfrac{x_{i} - \mu_i}{\sigma_i}
\end{equation}
Here, $\mu_i$ and $\sigma_i$ are the mean and standard deviation of the $i$-th feature, where $ i = \left\lbrace 1, \cdots, D \right\rbrace  $. 

\begin{figure}
\centering
\resizebox{0.885\textwidth}{!}{%
	\begin{tikzpicture}[>=Triangle, ->, thick, draw=black!50,
	        node distance = 6mm and 12mm,
	          start chain = going below,
	every pin edge/.style = {<-,shorten <=1pt},
	        neuron/.style = {circle, fill=#1,
	                         minimum size=17pt, inner sep=0pt},
	        annot/.style = {text width=4em, align=center}
	                        ]

	\foreach \y [count=\i] in {1,2,3,D}
	{
	\ifnum\i=3
	    \node[neuron=none, on chain] (I-\i)  {\Large $\vdots$};
	    \node[neuron=none, right=of I-3] (H1-\i)  {\Large $\vdots$};
	    \node[neuron=none, right=of H1-3] (H2-\i)  {\Large $\vdots$};
	    \node[neuron=none, right=of H2-3] (H3-\i)  {\Large $\vdots$};
	    \node[neuron=none, right=of H3-3] (H4-\i)  {\Large $\vdots$};
	\else
	    \node[neuron=green!50, on chain, pin=180: \Large Input \#\y] (I-\i)  {\Large $x_{\y}$};
	    \node[neuron=Purple!80, right=of I-\i] (H1-\i)  {\Large};
	    \node[neuron=Purple!80, right=of H1-\i] (H2-\i)  {\Large};
	    \node[neuron=Purple!80, right=of H2-\i] (H3-\i)  {\Large};
	    \node[neuron=Purple!80, right=of H3-\i] (H4-\i)  {\Large};
	\fi
	}

	\node[neuron=red!50,right=of H4-2]  (H5-1)   {\Large};
	\node[neuron=red!50,right=of H4-3]  (H5-2)   {\Large};
	\node[neuron=white,right=of H5-1]  (H6-1)   {\Large};
	\node[neuron=white,right=of H5-2]  (H6-2)   {\Large};

	\foreach \i in {1,2,4} \foreach \j in {1,2,4} { \draw (I-\i) edge (H1-\j);}
	  
	\foreach \i in {1,2,4} \foreach \j in {1,2,4} { \draw (H1-\i) edge (H2-\j);}  
	
	\foreach \i in {1,2,4} \foreach \j in {1,2,4} { \draw (H2-\i) edge (H3-\j);}  
	
	\foreach \i in {1,2,4} \foreach \j in {1,2,4} { \draw (H3-\i) edge (H4-\j);}  
	  
	\foreach \i in {1,2,4} { \draw (H4-\i) edge (H5-1);}   
	\foreach \i in {1,2,4} { \draw (H4-\i) edge (H5-2);}    
	  
	\draw (H5-1) edge (10,-1.21);  
	\draw (H5-2) edge (10,-2.42); 
	
	\node at (11,-1.8) {\Large Copresence};

	\node[annot,below=of I-4.center]        {\Large $ D $};
	\node[annot,below=of H1-4.center]        {\Large 500};
	\node[annot,below=of H2-4.center]        {\Large 300};
	\node[annot,below=of H3-4.center]        {\Large 100};
	\node[annot,below=of H4-4.center]        {\Large 20};
	\node[annot,below=of H5-1 |- H4-4.center] {\Large 2};
    \end{tikzpicture}
}%
    \caption{Structure of a neural network with four hidden layers (number of neurons is shown below) used in \name; the network takes as input the processed \gls{csi} data and outputs if two devices are copresent or not.}
    \label{fig:net-struct}
\end{figure}
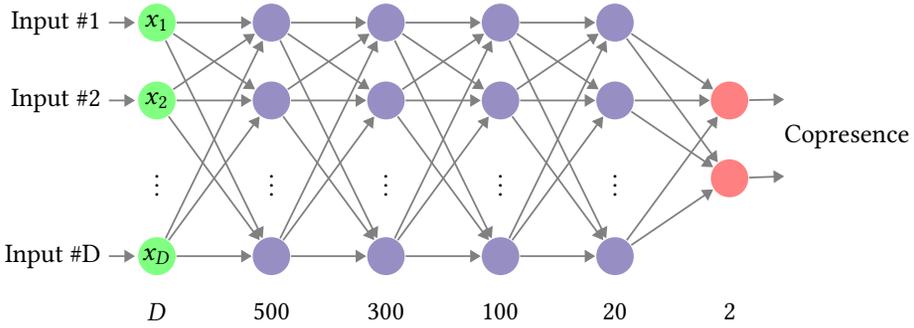

As discussed in~\autoref{subsec:sd-ml}, we choose neural networks as a machine learning classifier, allowing us to avoid manual feature engineering.
With the 5-fold \gls{cv}, we need to train a neural network five times, thus we design it with performance and reasonable training time in mind. 
Specifically, we implement a neural network in \textit{Keras}~\cite{Keras:2020} using four hidden layers with a decreasing number of neurons in each subsequent layer (i.e., 500, 300, 100, 20) and a softmax output layer (cf.~\autoref{fig:net-struct}).  
We denote each input vector as $ \mathbf{x} = \left[ x_1, \cdots, x_D \right] \in \mathbb{R}^{D \times 1} $, containing $ K' $ magnitude and $ K' $  phase values for every \gls{csi} measurement. 
Our network architecture has many neurons in the layers close to the input, reducing the number of neurons closer to the output. 
This gives the network enough capacity to learn a feature representation and gradually reduces the classifier's complexity in the last layers.
We do not use convolutions, as they introduce interdependencies among the features given the mask size and stride hyper-parameters.
To avoid this hyper-parameter search, we use a dense neural network with enough neurons and layers to model \gls{csi} data collected in different environments (e.g., office rooms, cars).
However, we leave as future work the test of pruning methods to reduce the network size as well as the behavior of convolutions and attention mechanisms~\cite{frankle2018lottery, lecun1989backpropagation, vaswani2017attention}.
We use the Leaky-ReLU~\cite{xu2015empirical} activation function for the hidden layers' neurons, as it avoids the vanishing gradient problem and also back-propagates gradients for negative values. 
For the loss function, we utilize cross-entropy.
To avoid overfitting, we keep the size of our network small and regularize by applying the dropout~\cite{srivastava2014dropout} after each hidden layer; the dropout rate is set to 0.2.
We use the Adam optimizer~\cite{kingma2014adam} with a learning rate of 0.001, and the batch size of 32, as often recommended~\cite{Brownlee:2020}.
We train the network until convergence, setting the number of epochs to 35 and 25 for 2.4 GHz and 5 GHz bands, respectively. 

\subsection{\name Prototype}
\label{subsec:prot}
We implement a \textit{Next2You} prototype working in real-time by combining the functionality of our \gls{csi} collector with the \textit{TensorFlow Lite} framework~\cite{TfL:2020}, allowing us to use pretrained neural network models directly on smartphones. 
Specifically, the prototype extracts \gls{csi} from incoming frames, processes it to the required input format (i.e., removing null subcarriers, performing feature normalization), and feeds the processed \gls{csi} to the model, which, in turn, outputs a copresence prediction. 
For the prototype, we reuse the same neural network architecture described in~\autoref{subsec:ml-pipeline}. 
To leverage \gls{csi} temporality, we accumulate successive \gls{csi} measurements within a time window, generating a copresence prediction for each measurement and using a majority vote to obtain the final decision. 
We consider time windows of 5 and 10 seconds, which provide a fair trade-off between the speed and reliability of copresence detection, respectively~\cite{Truong:2014}.


\section{Evaluation}
\label{sec:eval}
In this section, we present a comprehensive evaluation of \name based on the real-world \gls{csi} data that we collect. 

\begin{figure}
\centering
\begin{subfigure}[b]{0.375\textwidth}
	\centering
   	\includegraphics[width=\textwidth]{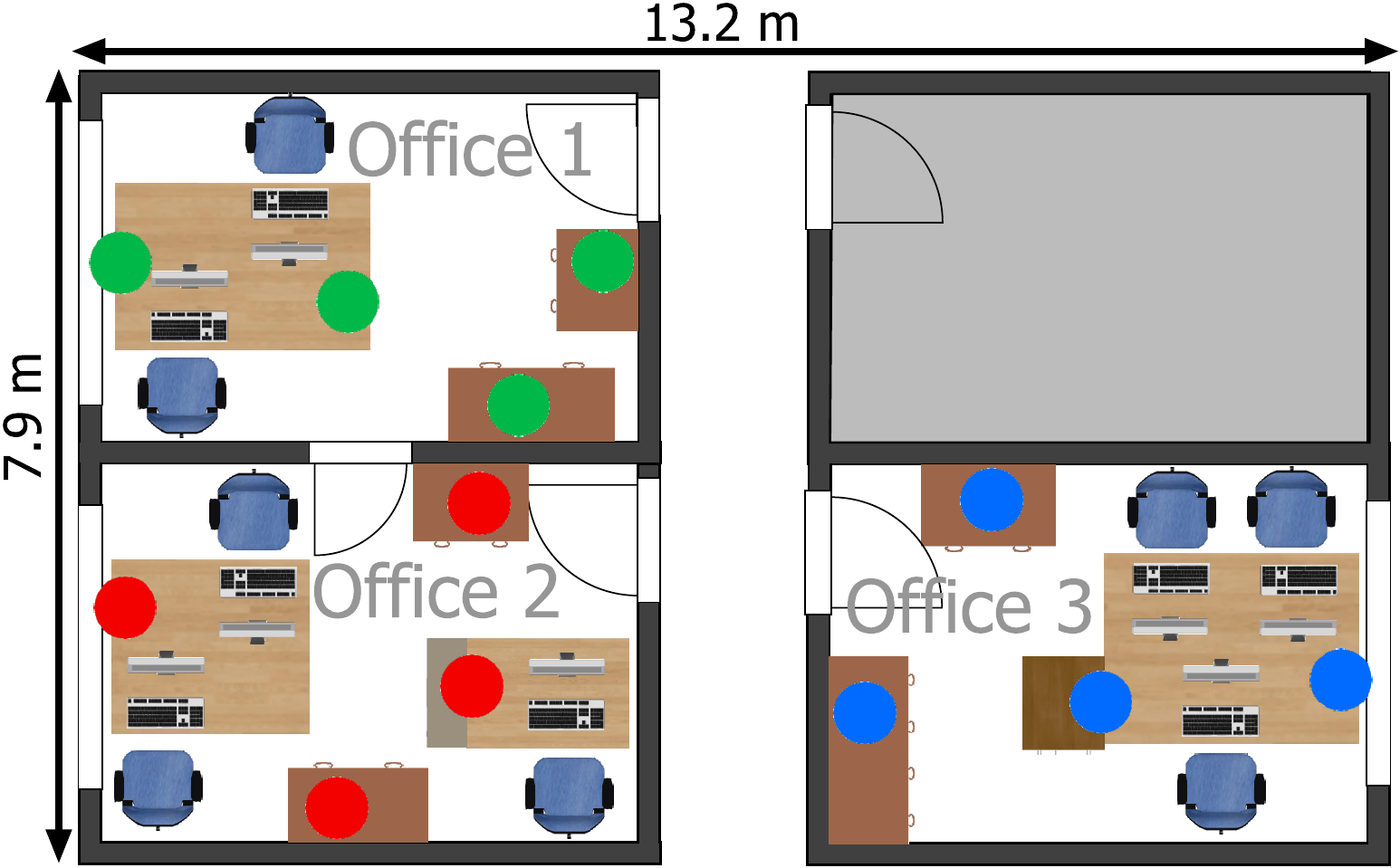}
   	\caption{Office}
   	\label{sf:setup-office}
\end{subfigure}
\begin{subfigure}[b]{0.375\textwidth}
	\centering
   	\includegraphics[width=\textwidth]{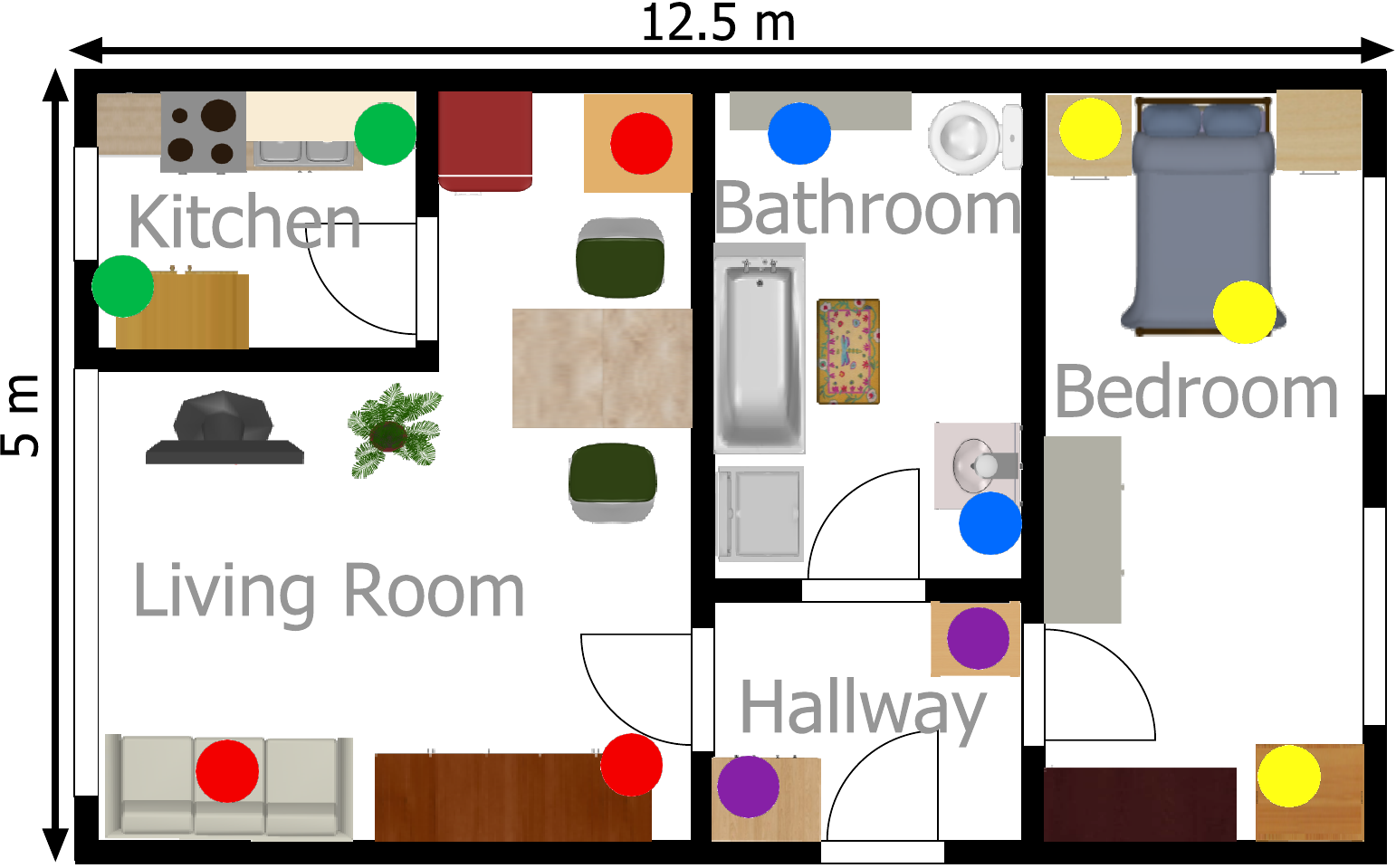}
   	\caption{Apartment}
    \label{sf:setup-flat}
\end{subfigure}
\begin{subfigure}[b]{0.375\textwidth}
	\centering
   	\includegraphics[width=\textwidth]{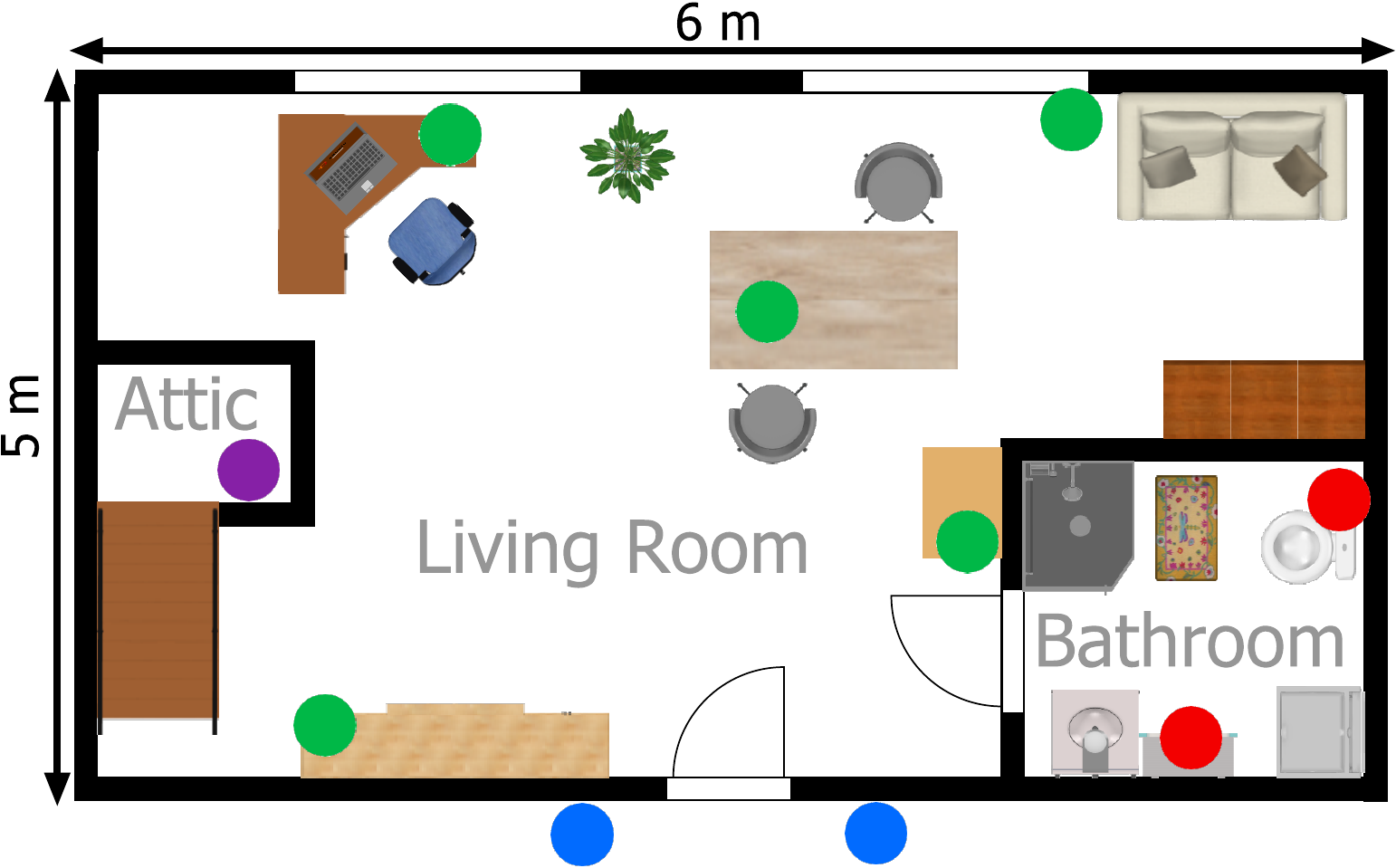}
   	\caption{House}
    \label{sf:setup-house}
\end{subfigure}
\begin{subfigure}[b]{0.375\textwidth}
	\centering
   	\includegraphics[width=\textwidth, cfbox=white 2pt 0pt]{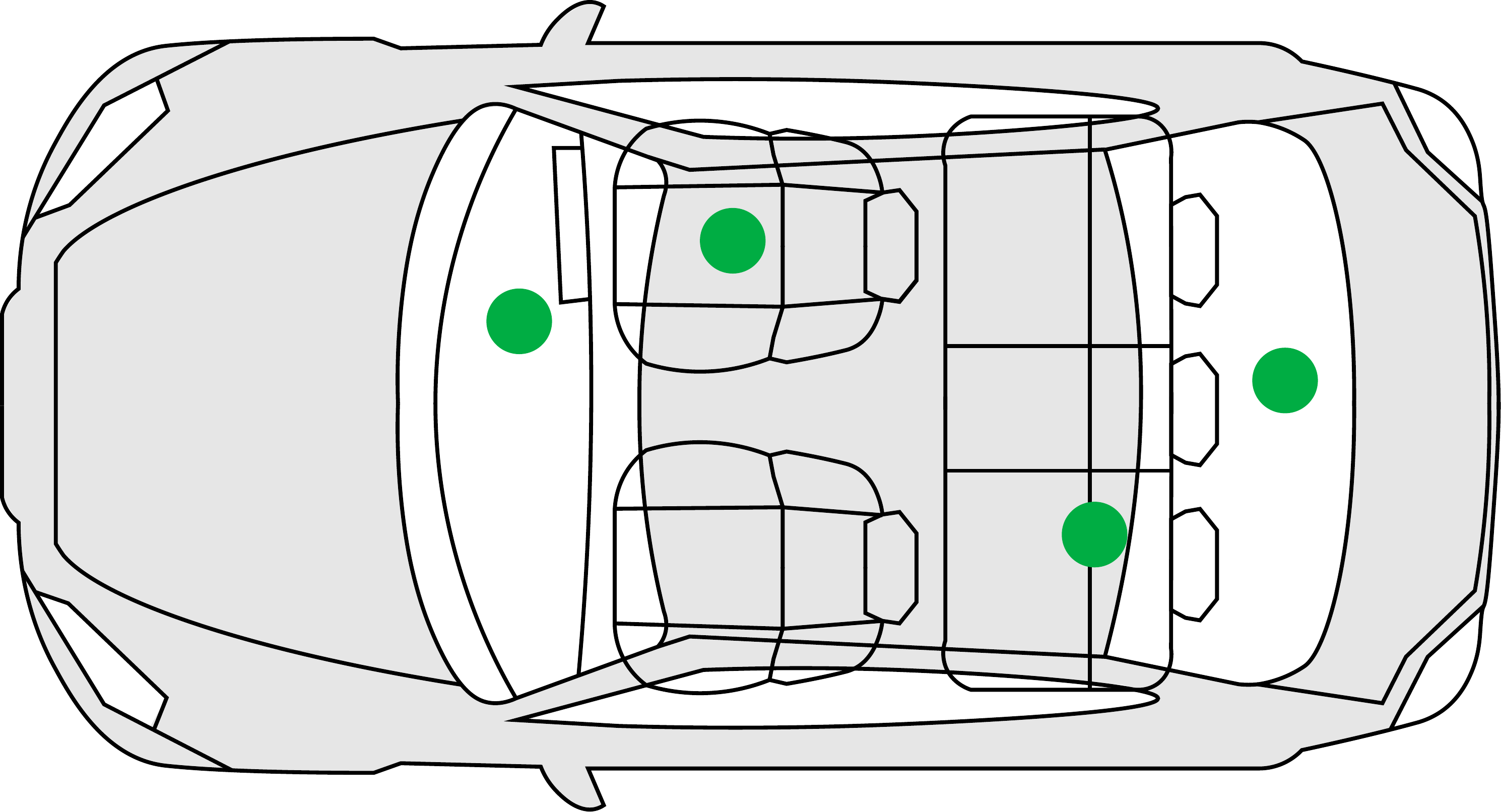}
   	\caption{Parked and moving cars}
    \label{sf:setup-car}
\end{subfigure}
\caption{Placement of smartphones collecting \gls{csi} data in different scenarios. The smartphones are marked with circles, which are of the same color for copresent devices. We depict the exterior and main obstacles in each scenario. The goal of \name is to distinguish between copresent and non-copresent devices located in different offices, rooms, or cars (e.g., devices in Office 1 from devices in Offices 2 and 3, cf.~\autoref{sf:setup-office}).}
\label{fig:exp-setup}
\end{figure}

\begin{table}
\small
\centering
	\caption{Overview of \gls{csi} data collection settings in different scenarios for 2.4 GHz and 5 GHz bands.}
	\label{tab:data-collect}
  \begin{tabular}{l|c|c|c|c}
  	\toprule
  	\multirow{2}{*}{Scenario} & \multirow{2}{*}{\makecell{Number of \\ devices}} & \multicolumn{2}{c|}{Accumulation round, minutes} & \multirow{2}{*}{\makecell{Overall time, \\ hours}} \\ 
  	& & \xspace \xspace \xspace \xspace 2.4 GHz \xspace \xspace \xspace \xspace & 5 GHz & \\
  	\midrule
  	Office{\large $^\dagger$} & 12 & 35 (20) & 25 (20) & 44 \\
  	-- Heterogeneous{\large $^*$} & 12 & 20 & 20 & 8 \\
  	-- Frame{\large $^*$} & 12 & 20 & 20 & 8 \\
  	-- Power{\large $^*$} & 12 & 20 & 20 & 8 \\
  	Apartment & 12 & 20 & 20 & 8 \\
  	House & 10 & 20 & 20 & 4.7 \\
  	Parked Cars & 8 & 10 & 10 & 2.7 \\
  	Moving Cars & 8 & 20 & 10 & 4 \\
  	\bottomrule
  \end{tabular}
  \smallskip\centering
  \center{{\large $^\dagger$}\small Collected over several days; {\large $^*$} \small Uses setup in~\autoref{sf:setup-office}; $()$ - Accumulation round at nighttime.}
\end{table}

\subsection{Experiment Setup}
\label{subsec:exp-setup}
To evaluate the capability of \name to detect copresence inside the same office, room, or car, we collect \gls{csi} data in five different scenarios: \textit{office}, \textit{apartment}, \textit{house} as well as \textit{parked} and \textit{moving cars}.  
These scenarios vary in terms of size and geometry, wall and obstacle materials, and the number of occupants.
Specifically, the \textit{office} consists of three office rooms: two adjacent offices and one across a hallway, occupied by one to three persons, the \textit{apartment} is a two-room flat inhabited by two people, while the \textit{house} is a single-person household, and the \textit{cars} are either parked side by side without any occupants in them, or being driven one after another with a single person inside each car for a total of 120~km.
To evaluate the impact of (1) heterogeneous devices, (2) different frame types, and (3) varying transmission power on copresence detection performance of \name, we collect \gls{csi} data in these three configurations reusing the office setup, resulting in \textit{heterogeneous}, \textit{frame}, and \textit{power} scenarios. 
In the \textit{office}, we collect the data over the course of several days, allowing us to assess copresence detection performance of \name at different times of the day (e.g., morning vs. night), while in other scenarios we record \gls{csi} for up to eight hours, as shown in~\autoref{tab:data-collect}. 
To capture \gls{csi} data, we use our app installed on Nexus 5 and Nexus 6P smartphones, and the Raspberry Pi 3 Model B+ (cf.~\autoref{subsec:data-app}), which are deployed in spots where \gls{iot} devices are typically found such as on a desk in the \textit{office}, on a dashboard in the \textit{car}, and near a TV set in the \textit{apartment} (cf.~\autoref{fig:exp-setup}).

To obtain representative \gls{csi} data, we collect it at each spot in the scenario.  
Specifically, during a data accumulation round, we have one device (\textit{verifier}) capturing \gls{csi} data from all other devices (\textit{provers}). 
For example, if a device on a window sill in Office 1 is a verifier, then it collects \gls{csi} data from three copresent provers in Office 1 and eight non-copresent provers located in Offices 2 and 3 (cf.~\autoref{sf:setup-office}). 
In the next data accumulation round, we change the position of the verifier (e.g., to a desk in Office 1), capturing \gls{csi} from all provers again, repeating this procedure until we obtain \gls{csi} data from each spot in the scenario (e.g., all circles in~\autoref{sf:setup-office}). 
We set the length of the data accumulation round based on the amount of \gls{csi} data available from 2.4 GHz and 5 GHz frequency bands (i.e., 112 and 484 magnitude and phase values, respectively) and complexity of the scenario (i.e., the amount of motion and number of obstacles), as presented in~\autoref{tab:data-collect}.
\\
\textbf{Reproducibility and Reusability.}
In total, we collect over 95 hours of \gls{csi} data. 
We release the collected  dataset and source code for data collection, evaluation, and \name prototype~\cite{Dataset:2021}. 
\\
\textbf{Ethical Considerations.}
Due to the sensitivity of the \gls{csi} data~\cite{Zhu:2020}, we obtained the approval for this study from our institutional ethical review board, the participants residing in experimental locations gave informed consent for the collection, use, and release of the \gls{csi} data.
\\
\textbf{Performance Metrics.}
\label{subsubsec:perf-metrics}
We evaluate whether \name can correctly classify copresent and non-copresent devices (e.g., distinguish devices in Office 1 from devices in Offices 2 and 3, cf.~\autoref{sf:setup-office}).
Specifically, we train a neural network model on the \gls{csi} data and compare its predictions with the ground truth, computing two performance metrics: \textit{\gls{auc}} and \textit{\gls{eer}}.
~The former shows how well the model can distinguish between copresent and non-copresent classes, with a higher \gls{auc} indicating a more accurately discriminative model; \gls{auc} is invariant to class imbalance compared to other metrics (e.g., accuracy). 
The latter is the intersection point of \textit{\gls{far}} and \textit{\gls{frr}}, balancing the number of misclassified non-copresent and copresent devices, respectively.  
The \gls{far} represents the \textit{security} of the system (i.e., non-copresent devices classified as copresent), while the \gls{frr} shows its \textit{usability} (i.e., copresent devices classified as non-copresent), thus a low \gls{eer} is desirable to achieve both these properties.

\subsection{Copresence Detection Performance}
\label{subsec:non-adv}
The results of \name copresence detection performance from our experiments are provided in~\autoref{tab:perf-nadv}, where
the two scenarios (i.e., \textit{frame} and \textit{power}) correspond to the active adversary, while the rest---represent the passive adversary (cf.~\autoref{sec:models}). 
In these experiments, we train a neural network on both copresent and non-copresent samples, investigating how well \name performs on unseen \gls{csi} data in advanced attack scenarios (cf.~\autoref{subsec:adv}). 
From~\autoref{tab:perf-nadv}, we observe that \name provides reliable copresence detection in different scenarios, achieving \glspl{eer} between 0 and 0.04.
\name shows more accurate copresence detection using the 5 GHz frequency band, and its performance decreases in larger environments with many obstacles as well as human and object motion (e.g., closing a door) compared to smaller stationary scenarios (e.g., \textit{office} vs. \textit{parked cars}).
\begin{table}
\small
\centering
	\caption{\gls{auc} and \gls{eer} of \name in different scenarios for 2.4 GHz and 5 GHz bands in the presence of  active and passive adversaries; the 5 GHz band shows slightly better results.}
	\label{tab:perf-nadv}
  \begin{tabular}{l|c|c|c|c|c}
  	\toprule
  	\multirow{2}{*}{Scenario} & \multirow{2}{*}{Adversary} & \multicolumn{2}{c|}{2.4 GHz} & \multicolumn{2}{c}{5 GHz} \\
  	& & \gls{auc} & \gls{eer} & \gls{auc} & \gls{eer} \\
  	\midrule
  	Office & Passive & 0.958 & 0.040 & 0.995 & 0.005 \\
  	-- Heterogeneous & Passive & 0.982 & 0.014 & 0.996 & 0.002 \\
  	-- Frame & Active & 0.988 & 0.010 & 0.993 & 0.005 \\
  	-- Power & Active & 0.995 & 0.002 & 0.999 & \colorbox{gray!50}{0.000} \\
  	Apartment & Passive & 0.961 & 0.022 & 0.984 & 0.012 \\
  	House & Passive & 0.993 & 0.007 & 0.984 & 0.015 \\
  	Parked Cars & Passive & 0.998 & \colorbox{gray!50}{0.001} & 0.999 & \colorbox{gray!50}{0.000} \\
  	Moving Cars & Passive & 0.996 & 0.002 & 0.997 & 0.001 \\
  	\bottomrule
  \end{tabular}
  \smallskip\centering
    \center{\small\colorbox{gray!50}{\parbox{.03\textwidth}{\transparent{0.0}\textcolor{gray!50}{0.00}}} - shows best achievable \gls{eer}.}
\end{table}
\\
\textbf{Impact of Frequency Band and Bandwidth.}
\label{subsub:fb-impact}
\autoref{tab:perf-nadv} reports that \gls{csi} data from the 5 GHz band allows distinguishing copresent and non-copresent devices more accurately.
We see three reasons for this result. 
First, the higher sensitivity of 5 GHz to path-loss causes more severe power attenuation from non-copresent devices,  limiting their communication range. 
Second, the shorter wavelength at 5 GHz (i.e., 6~cm), which is more easily perturbed by small-sized objects compared to 2.4 GHz (wavelength = 12.5~cm), allows discerning characteristics of the environment such as obstacle placement with higher granularity.
Third, the broader channel bandwidth at 5 GHz (i.e., 80 MHz) improves the time resolution of the \gls{cir}. 
Thus, more \gls{cir} paths can be distinguished from one another (cf.~\autoref{sf:cir_scenario}), resulting in more detailed \gls{csi} measurement, including information on distance ranges between devices inside the same environment. 

From a security perspective, the shorter range of 5 GHz is beneficial, as it forces the adversary to stay closer to legitimate devices. For example, in the \textit{moving cars} scenario, we had to drive two vehicles very slowly to maintain a distance of a few meters to be able to capture any \gls{csi} data from non-copresent devices.
In reality, following another car in such a way will immediately raise suspicion, imposing a physical barrier on the adversary's capability. 
However, the shorter communication range of 5 GHz can hinder usability---we see that the \gls{auc} and \gls{eer} for the \textit{apartment} and \textit{house} in 5 GHz are the lowest among the scenarios (cf.~\autoref{tab:perf-nadv}), indicating that copresent devices in larger rooms with many obstacles inside experience a rapid decrease in performance. 

\begin{figure}
\centering
\begin{subfigure}[b]{0.375\textwidth}
	\centering
   	\includegraphics[width=\textwidth]{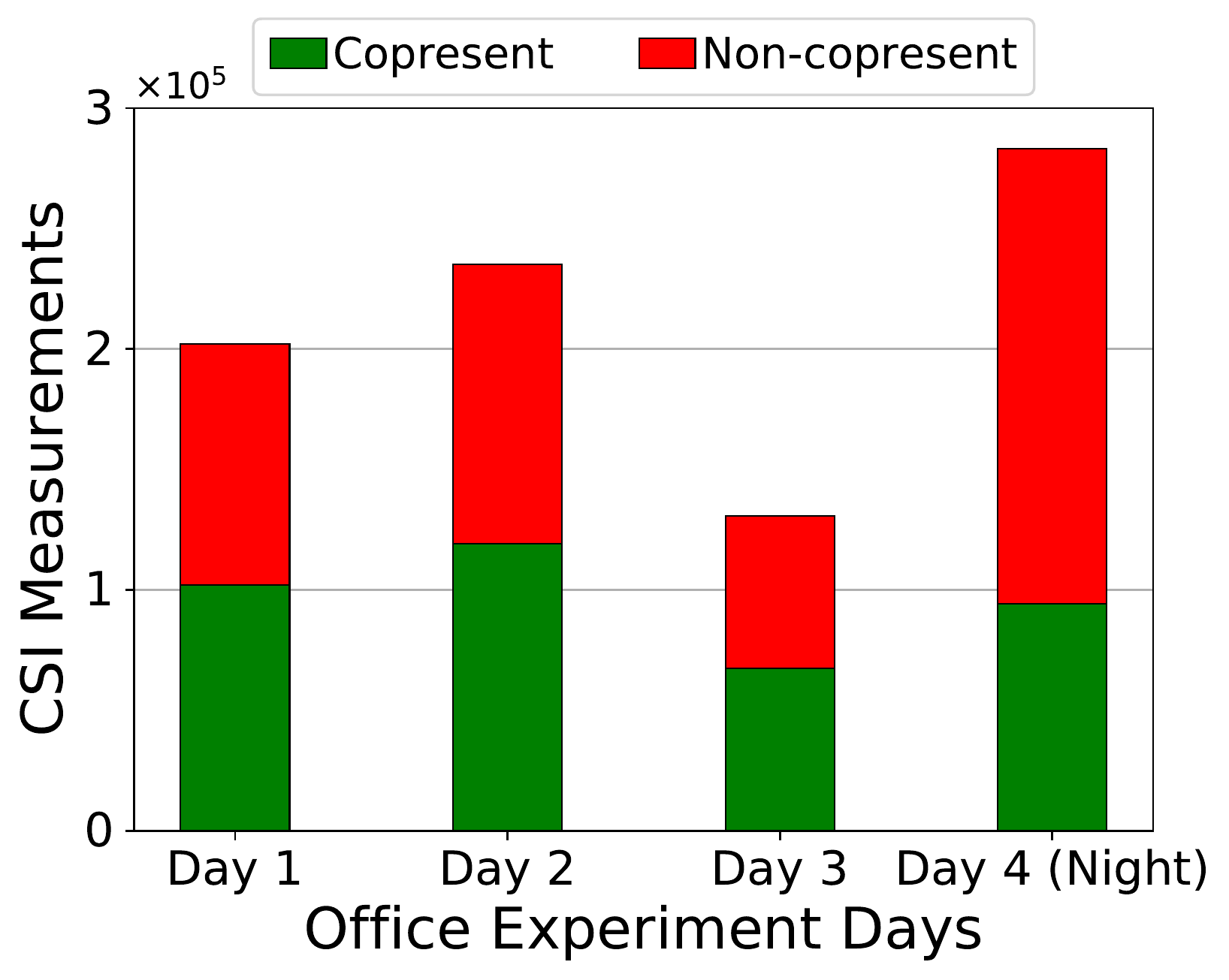}
   	\caption{2.4 GHz frequency band}
   	\label{sf:csi-meas-2.4}
\end{subfigure}
\begin{subfigure}[b]{0.375\textwidth}
	\centering
   	\includegraphics[width=\textwidth]{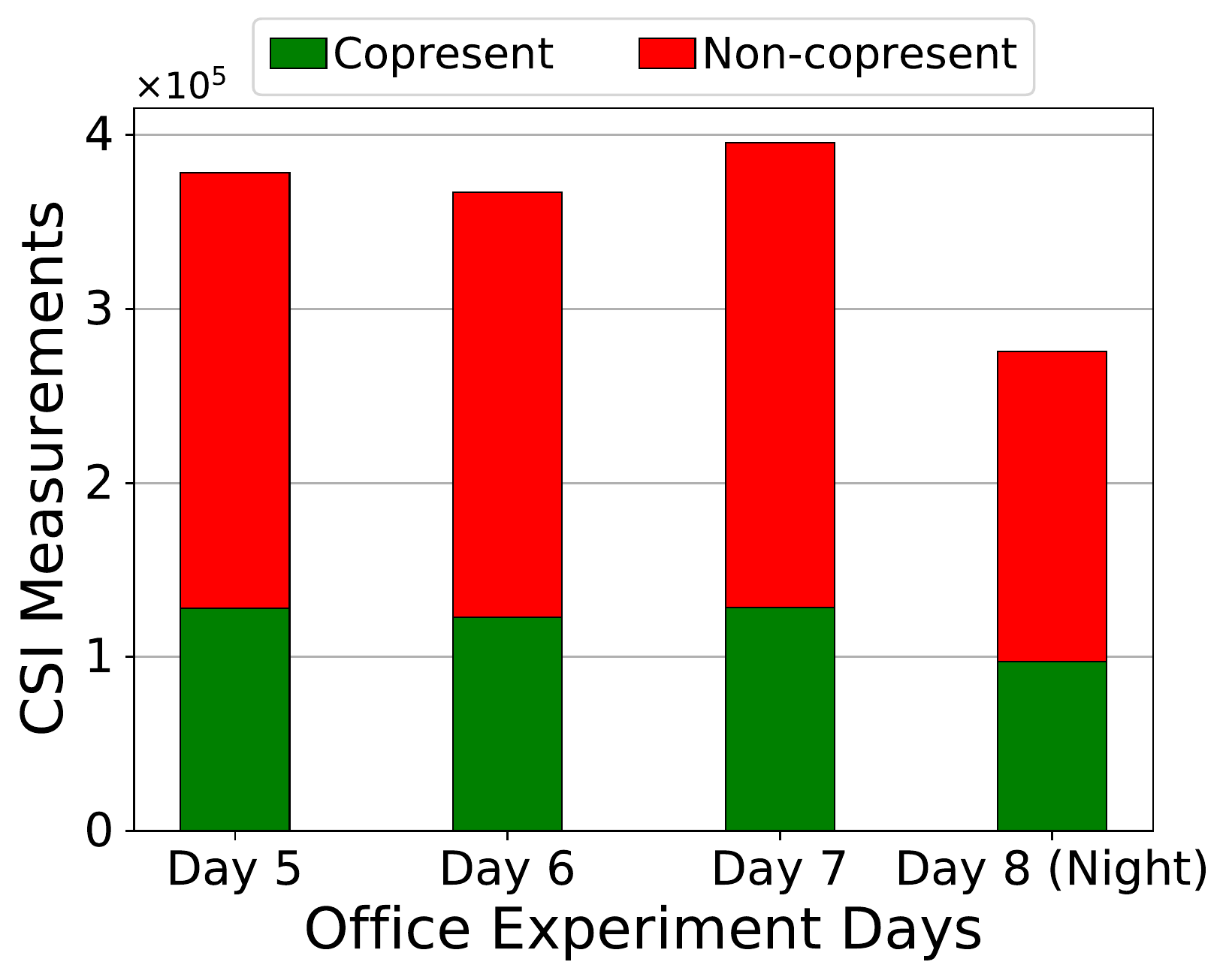}
   	\caption{5 GHz frequency band}
    \label{sf:csi-meas-5}
\end{subfigure}
\caption{Impact of frequency band on the number and ratio of \gls{csi} measurements collected by copresent and non-copresent devices in the office scenario on different days and at night; the number of \gls{csi} measurements collected at night is expected to be smaller due to a shorter data accumulation round (cf.~\autoref{tab:data-collect}).}
\label{fig:csi-meas}
\end{figure}

\begin{figure}
\centering
\begin{subfigure}[b]{0.375\textwidth}
	\centering
   	\includegraphics[width=\textwidth]{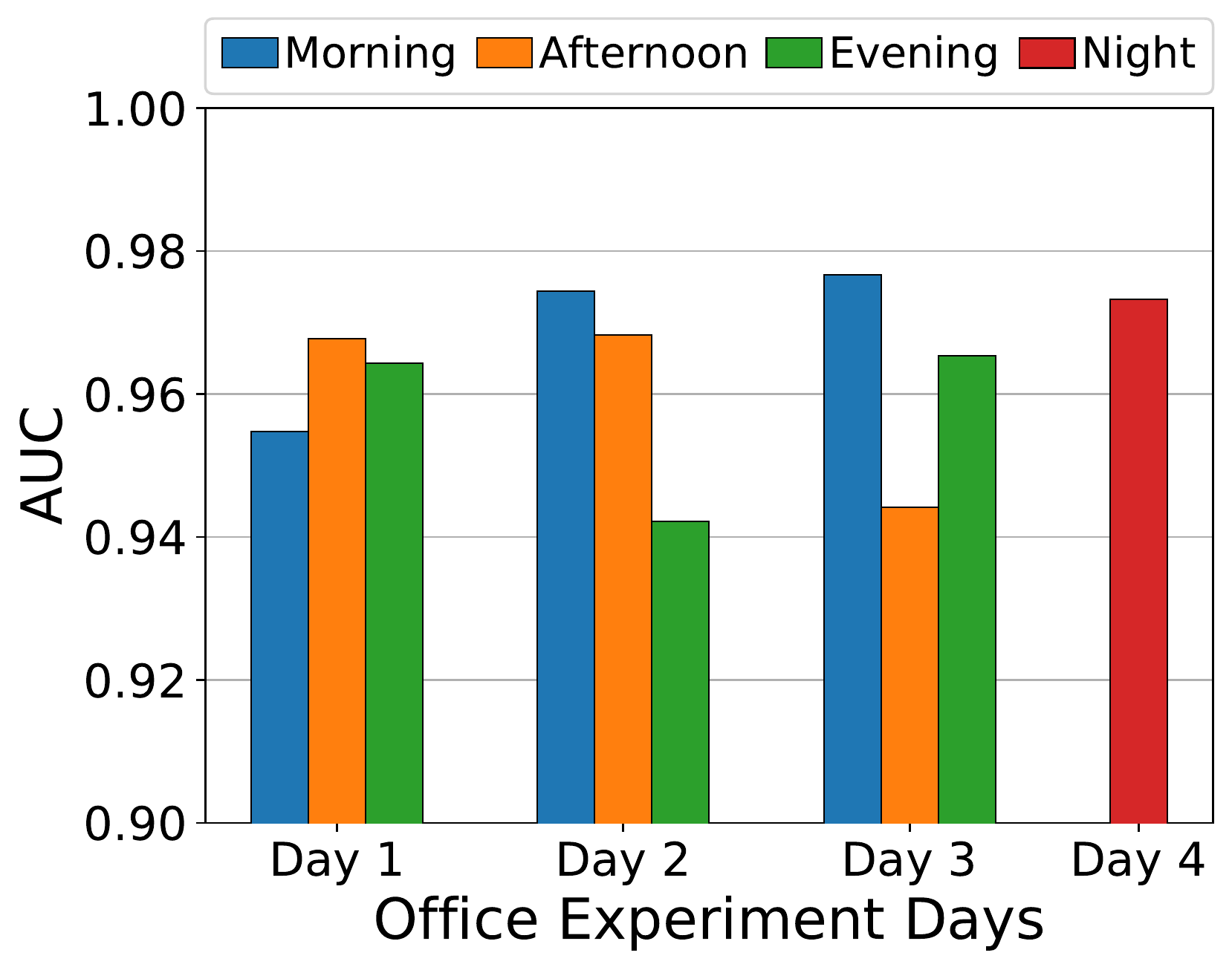}
   	\caption{2.4 GHz frequency band}
   	\label{sf:tod-2.4}
\end{subfigure}
\begin{subfigure}[b]{0.375\textwidth}
	\centering
   	\includegraphics[width=\textwidth]{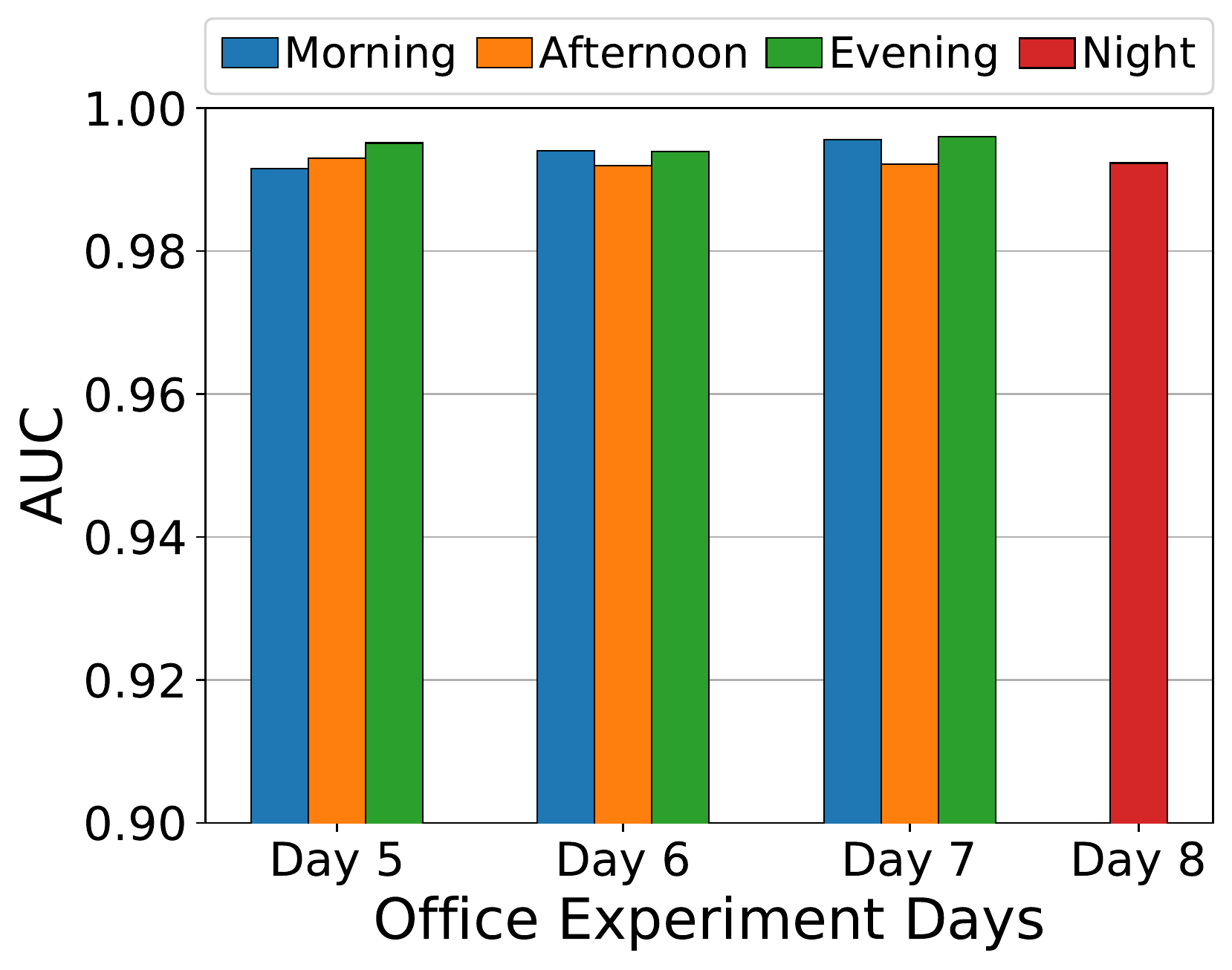}
   	\caption{5 GHz frequency band}
    \label{sf:tod-5}
\end{subfigure}
\caption{Impact of time of day on \gls{auc} of \name in the office scenario.}
\label{fig:tod-effect}
\end{figure}

We study the structure of our \gls{csi} data, finding that for the 2.4 GHz band in the \textit{office}, \textit{frame}, and \textit{heterogeneous} scenarios, the number of copresent and non-copresent \gls{csi} measurements does not conform to the expected ratio of roughly 30\% to 70\% (i.e., three copresent vs. eight non-copresent devices, cf.~\autoref{sf:setup-office}).
Instead, the number of copresent and non-copresent measurements is almost equal, which occurs only during working hours and not at night (cf.~\autoref{sf:csi-meas-2.4}).
In contrast, all 5 GHz measurements (cf.~\autoref{sf:csi-meas-5} for the \textit{office}) as well as 2.4 GHz in other scenarios, follow the expected ratio.
We attribute this behavior to the crowded spectrum in the 2.4 GHz band during working office hours (e.g., heavy Wi-Fi and Bluetooth traffic), preventing many frames from non-copresent \textit{provers} reaching the \textit{verifier} in their wireless range due to interference. 
The spectrum busyness generally reduces the overall number of collected \gls{csi} measurements (e.g., Day 3 vs. Day 4 (Night) in~\autoref{sf:csi-meas-2.4}).
This effect is beneficial for security, providing better separation between copresent and non-copresent devices.
However, it may hinder usability or availability of \name in large environments (e.g., lecture hall) with many wireless devices operating in the same spectrum. 
\\
\textbf{Impact of Time of Day.}
\autoref{fig:tod-effect} depicts the \gls{auc} performance of \name at different times of day (i.e., morning, afternoon, evening, and night) in the \textit{office} scenario. 
This result is obtained on the \gls{csi} data collected at the same time of day on different days following the procedure described in~\autoref{subsec:ml-pipeline}.
We see that during a day (i.e., morning till evening), the 5 GHz band shows both higher and more consistent \glspl{auc} compared to 2.4 GHz, confirming its suitability for small- and medium-sized environments.
In addition, \glspl{auc} of different days retain similar trends in both frequency bands, suggesting that such factors as spectrum busyness (e.g., possibly congested 2.4 GHz spectrum at Day 3, cf.~\autoref{sf:csi-meas-2.4}) and motion do not significantly affect the performance of \name.
This finding is confirmed by the \gls{auc} at night, which remains comparable to the day \glspl{auc} in both frequency bands, despite the absence of motion and minimum spectrum busyness. 
The high \gls{auc} at night shows that \name can cope with low-entropy context even for adjacent Offices 1 and 2 separated by a thin wall (cf.~\autoref{sf:setup-office}).
\\
\textbf{Impact of Heterogeneous Devices.}
\label{subsub:het-impact}
To evaluate the capability of \name to work on heterogeneous devices, we make a customized port of the Nexmon \gls{csi}-extractor~\cite{Schulz:2018} to the Nexus 6P smartphone, which has a \textit{different} Wi-Fi chipset than Nexus 5, allowing us to send frames and extract \gls{csi} data with Nexus 6P.
We find that due to two transmitting antennas in Nexus 6P, the \gls{csi} resulting from a frame sent by Nexus 6P differs\footnote{To the best of our knowledge, the official port of the Nexmon \gls{csi}-extractor on Nexus 6P now also supports single or multiple spacial streams, which should address this limitation (cf.~\url{https://github.com/seemoo-lab/nexmon_csi}).} from \gls{csi} extracted by Nexus 5, which contains a single antenna.
Hence, we use a Nexus 6P smartphone to capture \gls{csi} data from frames sent by Nexus 5 devices.
\autoref{tab:perf-nadv} shows that the \gls{auc} and \gls{eer} in the \textit{heterogeneous} scenario are moderately better compared to the \textit{office} (two scenarios share a setup, cf.~\autoref{sf:setup-office}) in both frequency bands, indicating the practicality of \name on heterogeneous devices.
To confirm this evidence, we use the Raspberry Pi 3 Model B+, which has a \textit{different} Wi-Fi chipset from Nexus smartphones, to collect \gls{csi} data from Nexus 5 devices in a setup similar to the \textit{heterogeneous} scenario (i.e., three office rooms with seven smartphones overall), obtaining even slightly higher \glspl{auc} of 0.9955 and 0.9988 for 2.4 GHz and 5 GHz bands, respectively.
\\
\textbf{Impact of Frame Type.}
To assess the robustness of \name, we change the transmitted data format extracting \gls{csi} from \textit{beacon} instead of \gls{qos} frames.
We do not observe significant changes in the structure of collected \gls{csi} data (e.g., similar trends for \gls{csi} measurements, cf.~\autoref{fig:csi-meas}).
The \gls{auc} and \gls{eer} in the \textit{frame} scenario are noticeably better compared to the \textit{office} in 2.4 GHz but are similar in the 5 GHz band (cf.~\autoref{tab:perf-nadv}).
We attribute the improved performance in 2.4 GHz to a shorter data accumulation round in the \textit{frame} scenario (i.e., 20 vs. 35 minutes in the \textit{office}, cf.~\autoref{tab:data-collect}), capturing fewer complex events such as relocation of obstacles, leading to more accurate copresence detection.
Thus, we demonstrate that \name is agnostic to the frame type, facilitating its deployability  (e.g., one verifier can execute \name with several provers, each using a different frame type). 
\\
\textbf{Impact of Transmission Power.}
\label{subsub:p-impact}
We study if an adversary can attack \name by increasing the transmission power of non-copresent devices to overcome  such effects as path-loss and shadowing~\cite{Shi:2017}.
To find by how much the power needs to be increased, we increment it until we obtain the expected ratio of \gls{csi} measurements (i.e., roughly 30\% to 70\%) for copresent and non-copresent devices at 2.4 GHz in the office setup during working hours. 
Thus, all frames sent by non-copresent provers reach the verifier despite the interference and blockages as if they are ``copresent''. 
We empirically find that the power needs to be increased by ten times to achieve this. 

\autoref{tab:perf-nadv} shows that the \gls{auc} and \gls{eer} of the \textit{power} scenario are significantly better compared to the \textit{office} in both frequency bands. 
We verify this result by finding no noticeable difference between \gls{csi} magnitude\footnote{Nexus devices have \gls{agc} enabled, thus power change is not reflected in the \gls{csi}, cf.~\url{https://github.com/seemoo-lab/mobisys2018_nexmon_channel_state_information_extractor/issues/2\#issuecomment-384088517}.} and its variance in both scenarios as well as normalizing the \gls{csi} magnitude by a unit power and obtaining unchanged \gls{auc} and \gls{eer}.
Based on the interpretation of hypotheses learned by our neural network from the \gls{csi} data (cf.~\autoref{sub:robust}), we conclude that increased power affects the statistical properties of \gls{csi}, making it easier to classify devices transmitting with different power.
Specifically, higher power produces additional \gls{cir} paths (cf.~\autoref{subsec:csi-rationale}) originated from the adversary's environment, distinguishing the \gls{csi} of non-copresent devices from copresent. 
\\
\textbf{Real-time Performance and Impact of Mobility.}
\label{subsub:n2y-prot}
We evaluate the real-time performance of \name in the \textit{office} scenario using our prototype (cf.~\autoref{subsec:prot}) in two cases.
First, we adjust the position of smartphones within 50~cm from their initial spots (cf. circles in~\autoref{sf:setup-office}), putting some devices on top of books or boxes. 
Then, a device at each spot in turn acts as the verifier predicting copresence in real-time, while all other devices are provers. 
In this case, \name correctly detects copresence around 95\% of the time in both frequency bands, regardless of the \gls{csi} measurement window (i.e., 5 or 10 seconds). 
Second, we challenge \name by introducing mobility. 
Specifically, the provers are deployed as shown in~\autoref{sf:setup-office}, whereas the verifier is carried by a user. 
The user continuously moves inside one of three offices or a hallway, approaching office doors from the hallway or walking close to a thin wall separating adjacent offices.
We note that the neural network models for 2.4 GHz and 5 GHz bands that are deployed on the mobile verifier have been trained on the \gls{csi} data collected by stationary devices, as described in~\autoref{subsec:exp-setup}. 

\begin{table}
\small
\centering
	\caption{\gls{far} and \gls{frr} of \name for 2.4 GHz and 5 GHz bands in the case of mobility.}
	\label{tab:perf-prot}
  \begin{tabular}{c|cc|cc|cc|cc}
  	\toprule
  	\multirow{3}{*}{Location} & \multicolumn{4}{c|}{2.4 GHz} & \multicolumn{4}{c}{5 GHz} \\
  	& \multicolumn{2}{c|}{5 sec} & \multicolumn{2}{c|}{10 sec} & \multicolumn{2}{c|}{5 sec} & \multicolumn{2}{c}{10 sec} \\
  	& \gls{far} & \gls{frr} & \gls{far} & \gls{frr} & \gls{far} & \gls{frr} & \gls{far} & \gls{frr} \\
  	\midrule
  	Hallway & 0.111 & n/a & 0.083 & n/a & 0.016 & n/a & 0.009 & n/a \\
  	Office 3 & 0.075 & 0.203 & 0.054 & 0.085 & 0.000 & 0.329 & 0.000 & 0.298 \\
  	Office 2 & 0.196 & 0.262 & 0.131 & 0.220 & 0.128 & 0.274 & 0.065 & 0.326 \\
  	Office 1 & 0.148 & 0.382 & 0.190 & 0.308 & 0.027 & 0.238 & 0.000 & 0.200 \\
  	\bottomrule
  \end{tabular}
\end{table}

\autoref{tab:perf-prot} shows the \gls{far} and \gls{frr} performance of \name in the case of mobility for each location, frequency band, and time window. 
We see that \gls{far} when the verifier moves in the hallway varies from 0.111 to 0.009, decreasing with the higher frequency band and longer time window. 
Exploring the misclassified provers, we find that \gls{far} is caused by devices that either have a line of sight with open office doors or located close to the hallway wall; closing offices' doors reduces \glspl{far} to almost zero for both frequency bands and time windows.
In Office 3, separated by the hallway, both \gls{far} and \gls{frr} steadily decrease with a longer time window for 2.4 GHz. 
Inspecting \gls{far} and \gls{frr}, we discover that the former is caused by devices from opposite Office 2 located near the door, while the latter comes from either heavily obstructed (i.e., window sill) or highest above the floor (i.e., office cabinet) devices. 
For 5 GHz, \gls{far} remains zero for both time windows, while \gls{frr} is relatively high and does not change significantly.
We find that a device on top of the office cabinet accounts for over 70\% of the \gls{frr}, which we attribute to heavier attenuation and reflections due to the moving verifier being blocked by the user as well as the least amount of training data available for devices high above the floor.
In adjacent Offices 1 and 2, \gls{far} and \gls{frr} follow the above trends but are higher, especially for 2.4 GHz, showing the impact of insufficiently separated environments; \gls{far} is overwhelmingly caused by devices from the adjacent office, while \gls{frr}, as before, comes from heavily obstructed and high above the floor devices.

The above results demonstrate that \name not only works reliably in real-time but also shows high potential for classifying mobile \gls{csi} data despite being trained on the stationary data. 
In other words, the neural network is able to make a generalization, providing fairly accurate copresence detection for device locations that were not in the training dataset. 
However, further research is required to improve this generalization capability, for example, by customizing the network's architecture to reduce the impact of user motion~\cite{abyaneh2018deep}. 
\\
\textbf{Comparison with Prior Work.}
\label{subsub:cmp}
We compare the \gls{eer} performance of \name with two state-of-the-art audio-based copresence detection schemes from Karapanos et al.~\cite{Karapanos:2015} and Truong et al.~\cite{Truong:2014}. These schemes are evaluated by Fomichev et al.~\cite{Fomichev:2019} in very similar to our \textit{office} and \textit{car} scenarios (i.e., adjacent offices, parked cars), allowing for a direct comparison. 
\name shows at least 16-times lower \glspl{eer} compared to the state-of-the-art audio schemes in the case of insufficiently separated environments such as adjacent offices, while in low-entropy context of parked cars \name preforms several orders of magnitude better (cf.~\autoref{tab:cmp-work}).

\subsection{Generalizability}
\label{sub:general}
We evaluate the capability of \name to generalize to new application scenarios (e.g., different apartment) by investigating how much effort is required for transfer learning, namely to reuse a pretrained neural network and retrain it with the data of the new environment. 
As stated in~\autoref{subsec:sd-ml}, we choose neural networks because of their inherent ability to automatically learn the feature representation.  
We leverage this property as follows: based on our network structure (cf.~\autoref{fig:net-struct}), we observe that the first two layers (i.e., 500 and 300 neurons, respectively) are mostly dedicated  to learning the \gls{csi} feature representation, while the last two layers (i.e., 100 and 20 neurons, respectively) are focusing on the classification task. 
Thus, we can learn the \gls{csi} feature representation once on the most comprehensive \textit{office} dataset that contains the largest amount of \gls{csi} data, capturing complex geometry and motion, and then reuse this network and retrain it on other scenarios, significantly reducing the computations required for the learning step.  

\begin{table}
\small
\centering
	\caption{\gls{eer} comparison between \name and state-of-the-art copresence detection schemes based on audio in the office and car scenarios (we present best achievable \glspl{eer} for each scheme).}
	\label{tab:cmp-work}
  \begin{tabular}{c|cc|cc}
  	\toprule
  	\multirow{2}{*}{Scheme} & \multicolumn{2}{c|}{\gls{eer} (Office)} & \multicolumn{2}{c}{\gls{eer} (Car)} \\
  	& Full{\large $^*$} & Night & Moving & Parked \\
  	\midrule
  	\name (this work) & 0.005 & 0.005 & 0.001 & 0.000 \\
  	Karapanos et al.~\cite{Karapanos:2015} & 0.098 & 0.090 & 0.006 & 0.037 \\
  	Truong et al.~\cite{Truong:2014} & 0.084 & 0.080 & 0.111 & 0.271 \\
  	\bottomrule
  \end{tabular}
    \smallskip\centering
  \center{{\large $^*$}\small Aggregated performance over several days, including all times of day (i.e., morning till night).}
\end{table}

\begin{table}
\small
\centering
	\caption{\gls{auc} obtained evaluating the generalizability of \name for 2.4 GHz and 5 GHz bands.}
	\label{tab:general}
  \begin{tabular}{l|cc}
  	\toprule
  	\multirow{2}{*}{Scenario} & \multicolumn{2}{c}{\gls{auc}} \\
  	& 2.4 GHz & 5 GHz \\
  	\midrule
  	Office & n/a & n/a \\
  	-- Heterogeneous & 0.981 & 0.995 \\
  	-- Frame & 0.986 &  0.985 \\
  	-- Power & 0.987 &  0.995 \\
  	Apartment & 0.964 &  0.982 \\
  	House & 0.986 & 0.976 \\
  	Parked Cars & 0.999 & 0.999 \\
  	Moving Cars & 0.997 & 0.998 \\
  	\bottomrule
  \end{tabular}
\end{table}

Specifically, we take the \textit{office} model, which can be efficiently trained on a powerful machine or in the cloud, and set the first two layers to be non-trainable. 
We then retrain the last two layers of such \textit{office} model on the train set of another scenario (e.g., \textit{house}) and use the obtained model to classify copresence on the test set of this scenario, performing this procedure for each of our scenarios except the \textit{office}.
The resulting \glspl{auc} are within one percentage point from the \glspl{auc} of scenario models trained from scratch (cf.~\autoref{tab:perf-nadv} vs.~\autoref{tab:general}), confirming that our network architecture allows for feature representation and transfer learning successfully. 
Since the retrained model is much simpler, we can train for fewer epochs (e.g., we use 10) and reduce the number of \gls{flops} in the forward training loop of our neural network by a factor of 7 and 13 for 2.4 GHz and 5 GHz bands, respectively. 
Given that modern end-user devices already perform in the giga\gls{flops}~\cite{FLOPs:2020} range, we consider it feasible to deploy and retrain \name in new scenarios using the described approach.

As all context-based schemes utilizing machine learning, \name requires initial data collection when being deployed in a new environment~\cite{Wu:2020}. For stationary devices, \gls{csi} data can be collected in the background: given a typical rate of 3--5 packets per second on \gls{iot} devices~\cite{Zhu:2020}, a representative dataset can be obtained in a matter of minutes. For mobile devices, a user will need to walk inside the environment (e.g., a room) for this time to collect typical \gls{csi}, which can similarly be done in the background during user's daily routine (e.g., office hours). 

\subsection{Interpretability of \name Copresence Detection}	
\label{sub:robust}
Our results demonstrate that \name performs well in classifying copresent and non-copresent devices in a variety of scenarios (cf.~\autoref{tab:perf-nadv}), and it generalizes to new environments (cf.~\autoref{sub:general}).
However, neural networks might not learn the right hypothesis, yet achieve high classification results for the wrong reasons~\cite{lapuschkin2019unmasking}.
To avoid this, we need to understand which factors play a role in a copresence decision produced by \name. 
Since \gls{csi} captures the combined scattering, path-loss, shadowing, and multi-path effects~\cite{Shi:2017}, which cannot be easily discerned, we identify features in the \gls{csi} data contributing to copresence detection and validate our findings against prior work on \gls{csi}-based localization and identification.

To interpret the copresence decision-making of \name, we apply a recently introduced \textit{\gls{rrr}} method~\cite{Ross:2017}, which identifies relevant features used by a neural network in a classification process. 
Thus, we can not only quantify parts of the \gls{csi} determining copresence but also harden \name against attacks by combining multiple hypotheses learned from \gls{csi} data (cf.~\autoref{subsec:adv}).
We utilize the \gls{rrr} method as follows (details in~\autoref{sec:appx0}): (1) we train a neural network on our \gls{csi} data, obtaining a gradient mask, containing the features considered important by the network, (2) using this mask we find relative importance of these features in the whole training data, (3) we select and penalize features with relative importance above 10\%, retraining the network on the same data to make it learn another hypothesis, (4) we repeat the above steps (1)--(3), increasing the number of penalized features until the \gls{auc} on the test data drops below 0.85; we choose the 10\% feature importance and 0.85 stopping thresholds empirically.

\begin{figure}
\centering
\begin{subfigure}[b]{0.375\textwidth}
	\centering
   	\includegraphics[width=\textwidth]{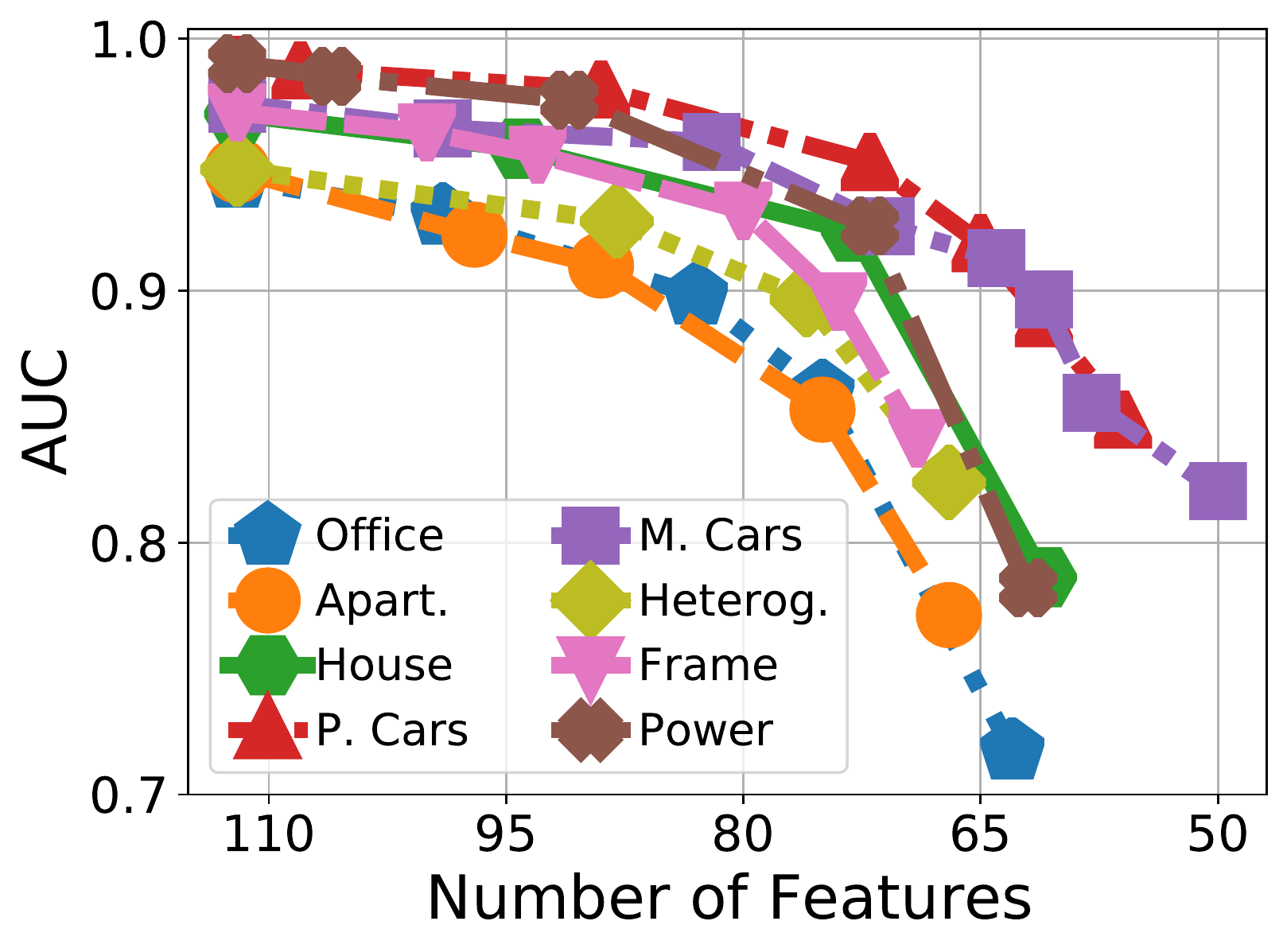}
   	\caption{2.4 GHz frequency band}
   	\label{sf:rrr-effect-2.4}
\end{subfigure}
\begin{subfigure}[b]{0.375\textwidth}
	\centering
   	\includegraphics[width=\textwidth]{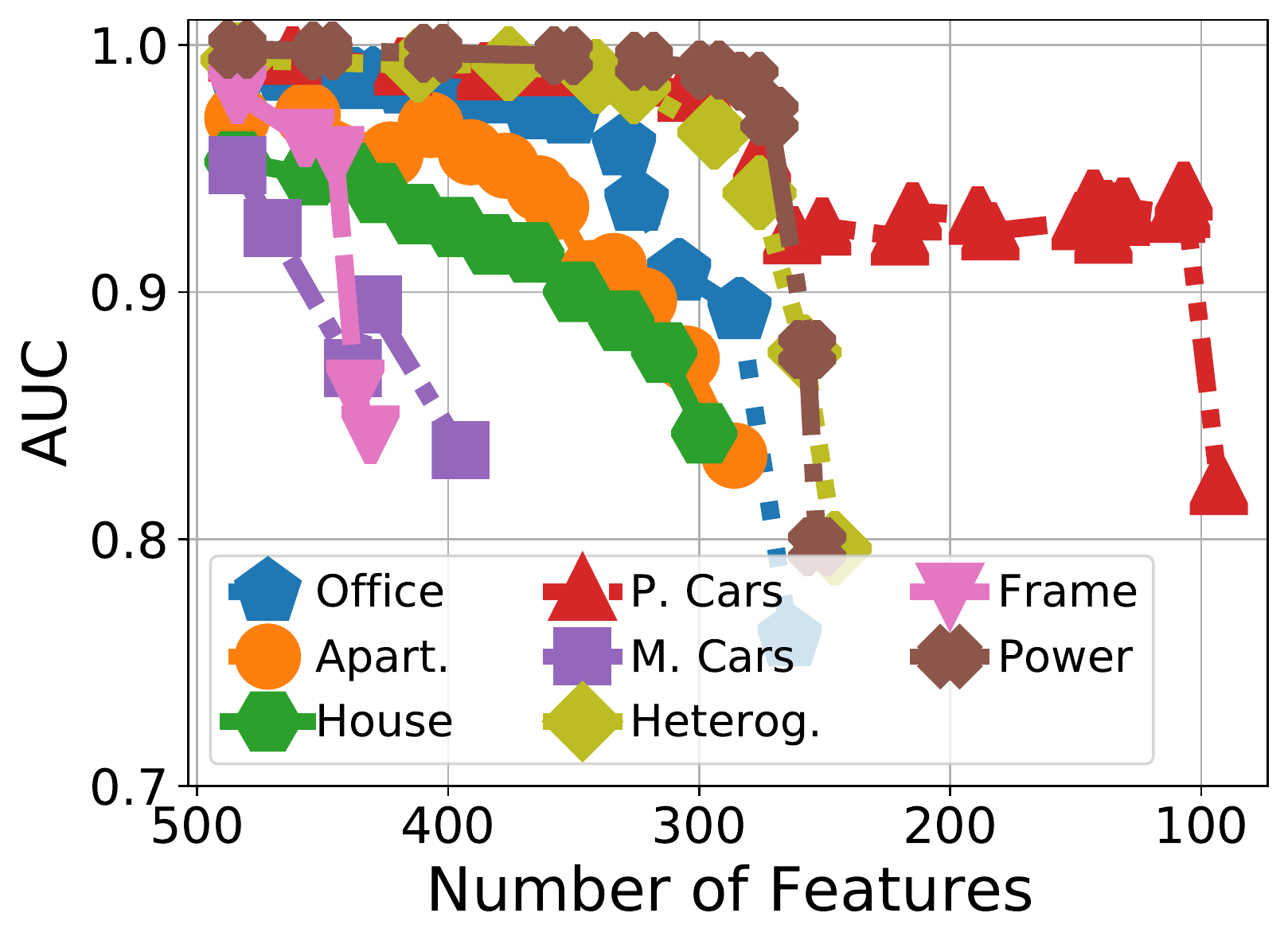}
   	\caption{5 GHz frequency band}
    \label{sf:rrr-effect-5}
\end{subfigure}
\caption{Right for the Right Reasons method applied to the \gls{csi} data of different scenarios. A marker denotes a neural network model trained on a different set of features, showing its \gls{auc} on the test data.}
\label{fig:rrr-effect}
\end{figure}

\autoref{fig:rrr-effect} shows the \gls{auc} performance of different neural network models produced by the \gls{rrr} method in 2.4 GHz and 5 GHz bands. 
We see that multiple models relying on different features can be trained from the \gls{csi} data, retaining high \gls{auc}, confirming the suitability of \gls{csi} for copresence detection. 
In larger and more complex scenarios (i.e., \textit{office}, \textit{apartment}, and \textit{house}) more features are required to classify copresent and non-copresent devices, while in the smaller and simpler \textit{parked cars}, we may need as few as four features. 
We observe distinct feature patterns in \textit{heterogeneous} and \textit{power} scenarios.
In the former scenario, more features are required to distinguish copresent and non-copresent devices compared to the \textit{office} (two scenarios share a setup, cf.~\autoref{sf:setup-office}) in both frequency bands. 
Thus, a neural network likely captures properties of \gls{csi}-extracting hardware, which may by used to fingerprint specific types of devices, providing an additional layer of protection. 
In the latter scenario, the models often rely on 6--8 important features, originating from the \gls{csi} of subcarriers uniformly spread over a Wi-Fi channel, indicating that increased power affects statistical properties of the \gls{csi}.

We find that the neural network models in~\autoref{fig:rrr-effect} rely on \gls{csi} magnitude for all scenarios except the \textit{parked} and \textit{moving cars}, which start to use phases once magnitude features are penalized.
In comparable scenarios (e.g., \textit{office} and \textit{house}), the models without penalization use similar important features resulting from the \gls{csi} of subcarriers located at the beginning, in the middle, or at the end of a Wi-Fi channel.
These subcarriers are relevant because of their stability and low susceptibility to noise~\cite{Shi:2017}.
Thus, they can accurately capture a particular effect in the environment such as shadowing due to large objects or scattering caused by window grids. 
Our findings about the important \gls{csi} features for copresence detection agree with prior work.
For example, the \gls{csi} magnitude is known to be relevant in complex scenarios with many obstacles and human motion~\cite{Gong:2016}, while the phases contain too much noise, and thus are not useful~\cite{Liu:2017}. 
In the environments with shorter distances and fewer obstacles between devices, the phases can be successfully utilized~\cite{Sen:2012}, as we see in the \textit{parked} and \textit{moving cars}. 
We find that the magnitude and phases of adjacent Wi-Fi subcarriers are jointly identified as important features because similar frequencies are likely to be affected by the same phenomenon (e.g., scattering)~\cite{Shi:2017}.
These results demonstrate the soundness of the \gls{rrr} interpretations, confirming that \name is capable of learning a robust wireless fingerprint of the environment embedded in the \gls{csi}.

\begin{figure}
\centering
\begin{subfigure}[b]{0.375\textwidth}
	\centering
   	\includegraphics[width=\textwidth]{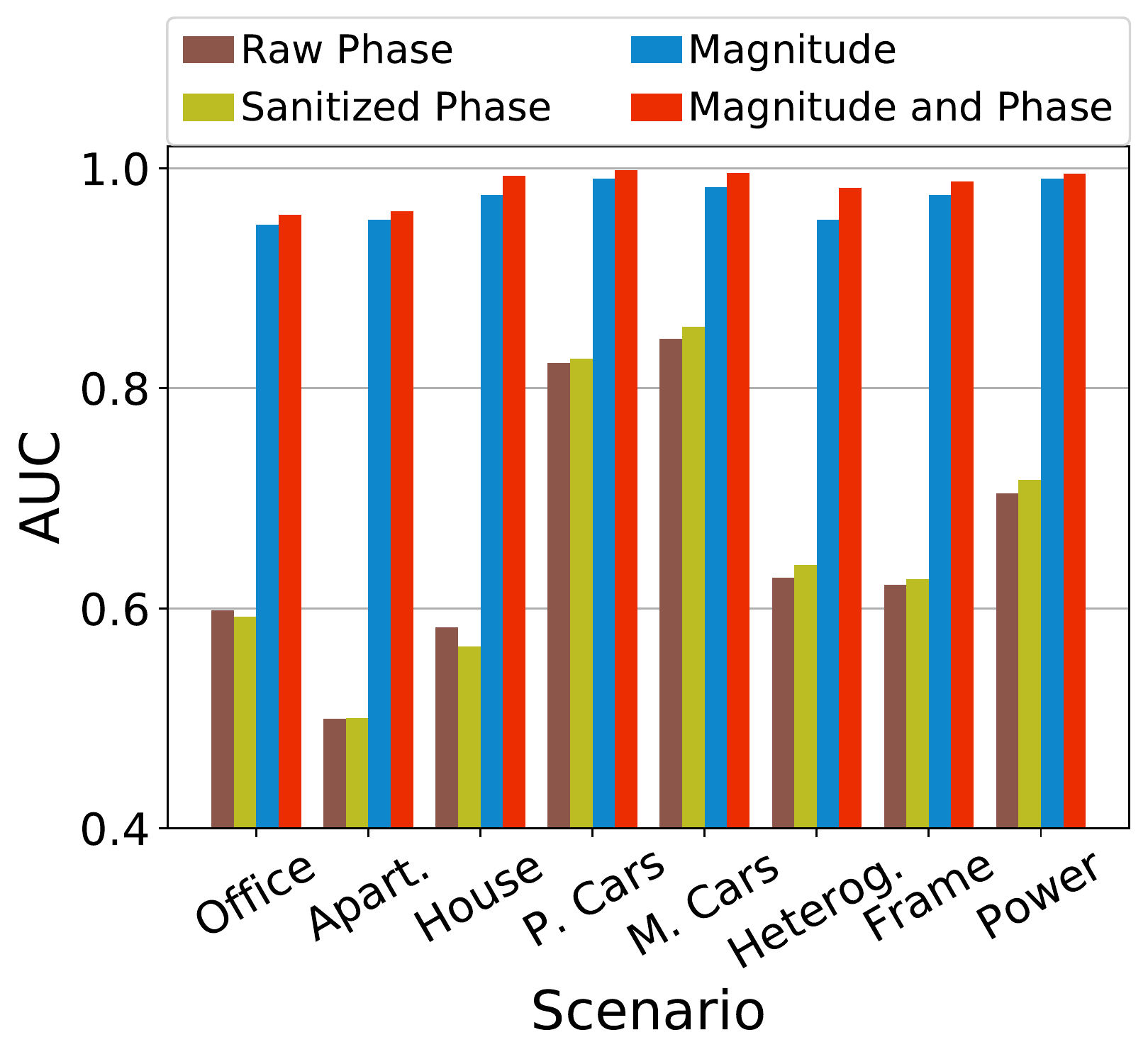}
   	\caption{2.4 GHz frequency band}
   	\label{sf:csi-impact-2.4}
\end{subfigure}
\begin{subfigure}[b]{0.375\textwidth}
	\centering
   	\includegraphics[width=\textwidth]{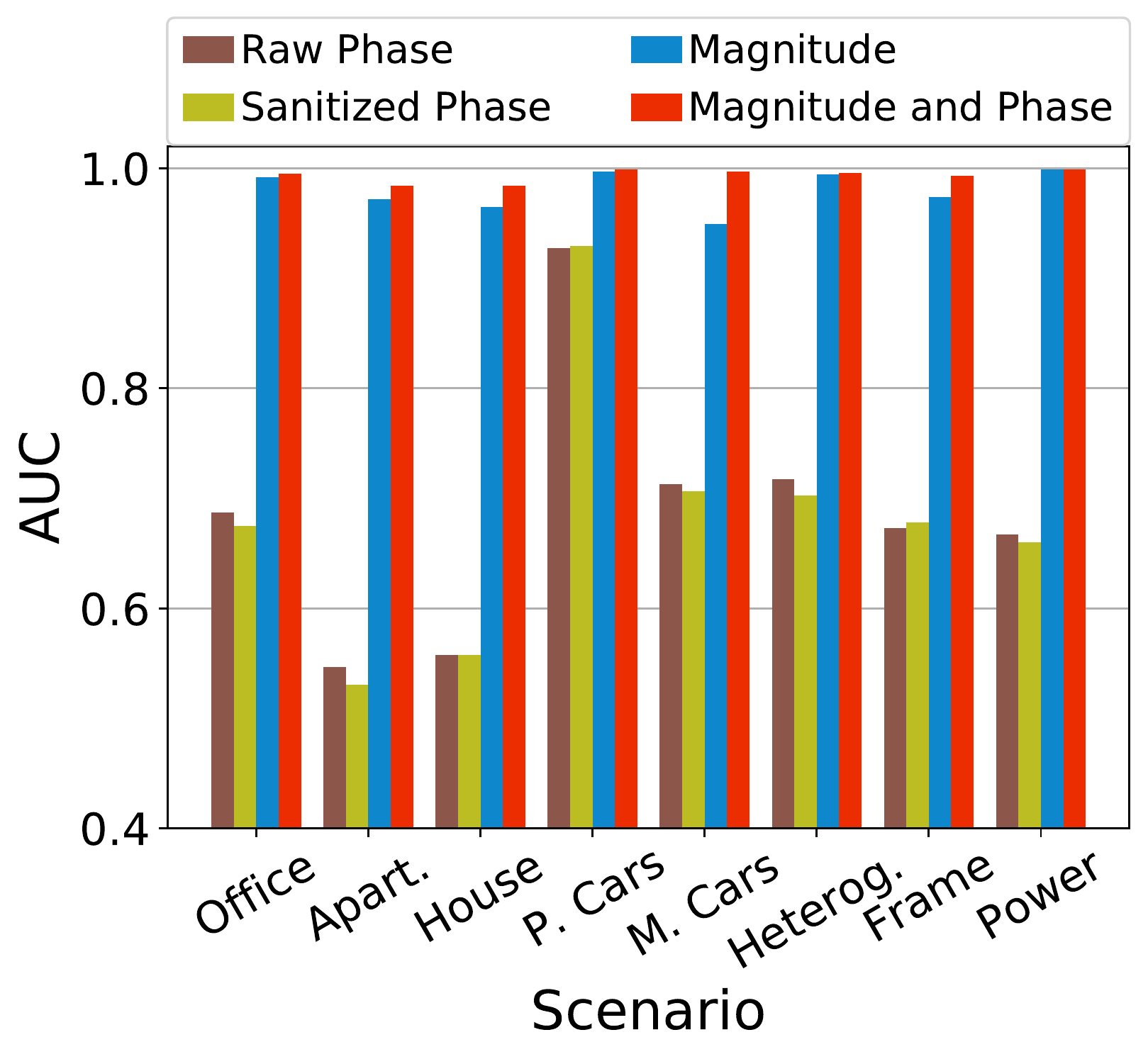}
   	\caption{5 GHz frequency band}
    \label{sf:csi-impact-5}
\end{subfigure}
\caption{\gls{auc} of \name when a neural network is trained on either phase (raw or sanitized) or magnitude features in different scenarios; we also show \gls{auc} when the network uses both magnitude and phase.}
\label{fig:csi-impact}
\end{figure}

To quantify the contributions of \gls{csi} features to \name classification performance, we train a neural network on either magnitudes or phases (cf.~\autoref{fig:csi-impact}). 
We apply a phase sanitization method by Sen et al.~\cite{Sen:2012} to overcome the problem of random phase behavior caused by unsynchronized clocks between devices and random noise, training the network on both raw and sanitized phases to allow their comparison.
We see that in complex scenarios (e.g., \textit{apartment}) copresence detection based on \gls{csi} phases is infeasible, reaching \glspl{auc} of around 0.5, indicating that the neural network classification is no better than a random guess. 
However, in the simpler environments of \textit{parked} and \textit{moving cars}, the phases become feasible for copresence detection, showing \glspl{auc} above 0.8 and 0.9 for 2.4 GHz and 5 GHz bands, respectively. 
The sanitized phases perform marginally better than the raw phases, which is in contrast with findings of prior work~\cite{Sen:2012, Qian:2014}. 
Such a discrepancy is due simpler hardware in our experiments (i.e., smartphones), containing a single antenna, which has lower sensitivity and higher susceptibility to noise, compared to multiple antennas in routers used by the prior work. 
Our phase classification results agree with previous research, showing the higher relevance of 5 GHz phases for localization~\cite{Wang:2017}.
The \gls{auc} performance based on magnitudes is stable across our scenarios, however using both magnitude and phase features results in consistently higher \glspl{auc} (cf.~\autoref{fig:csi-impact}).
Thus, \gls{csi} phases indeed contain relevant copresence information, which can be utilized by a neural network to improve the overall classification performance. 

The above results demonstrate the soundness of \name utilizing \gls{csi} and neural networks for copresence detection. 
We see a direct impact of the environment complexity on the capability of a neural network to make copresence predictions from the \gls{csi} data.
The fact that the neural network captures distinct properties of diverse \gls{csi} data (i.e., heterogeneity, transmission power) confirms its inherent ability for autonomous representation learning and suitability for \name.
We also find that using off-the-shelf devices in real-world scenarios may render existing methods for leveraging \gls{csi} inefficient (e.g., phase sanitization), urging the need to conduct experiments in realistic setups with heterogeneous hardware. 

\subsection{Advanced Attack Scenarios}
\label{subsec:adv}
We investigate the robustness of \name to advanced attacks and propose mitigation strategies. 
In the first attack, the adversary either precollects \gls{csi} data in a similar environment (e.g., car), where legitimate devices execute \name, or they train a neural network model on a comprehensive dataset (e.g., \textit{office}) and use this model to predict copresence on the \gls{csi} data of simpler scenarios (e.g., \textit{parked cars}).  
We find that such threats are feasible allowing the adversary to predict copresence with relatively high \glspl{auc} between 0.8 and 0.9. 
To mitigate this attack, we utilize multiple neural network models, each learning a different hypothesis, generated by the \gls{rrr} method (cf.~\autoref{sub:robust}). 

\begin{figure}
\captionsetup[subfigure]{justification=centering}
\centering
\begin{subfigure}[b]{0.375\textwidth}
	\centering
   	\includegraphics[width=\textwidth]{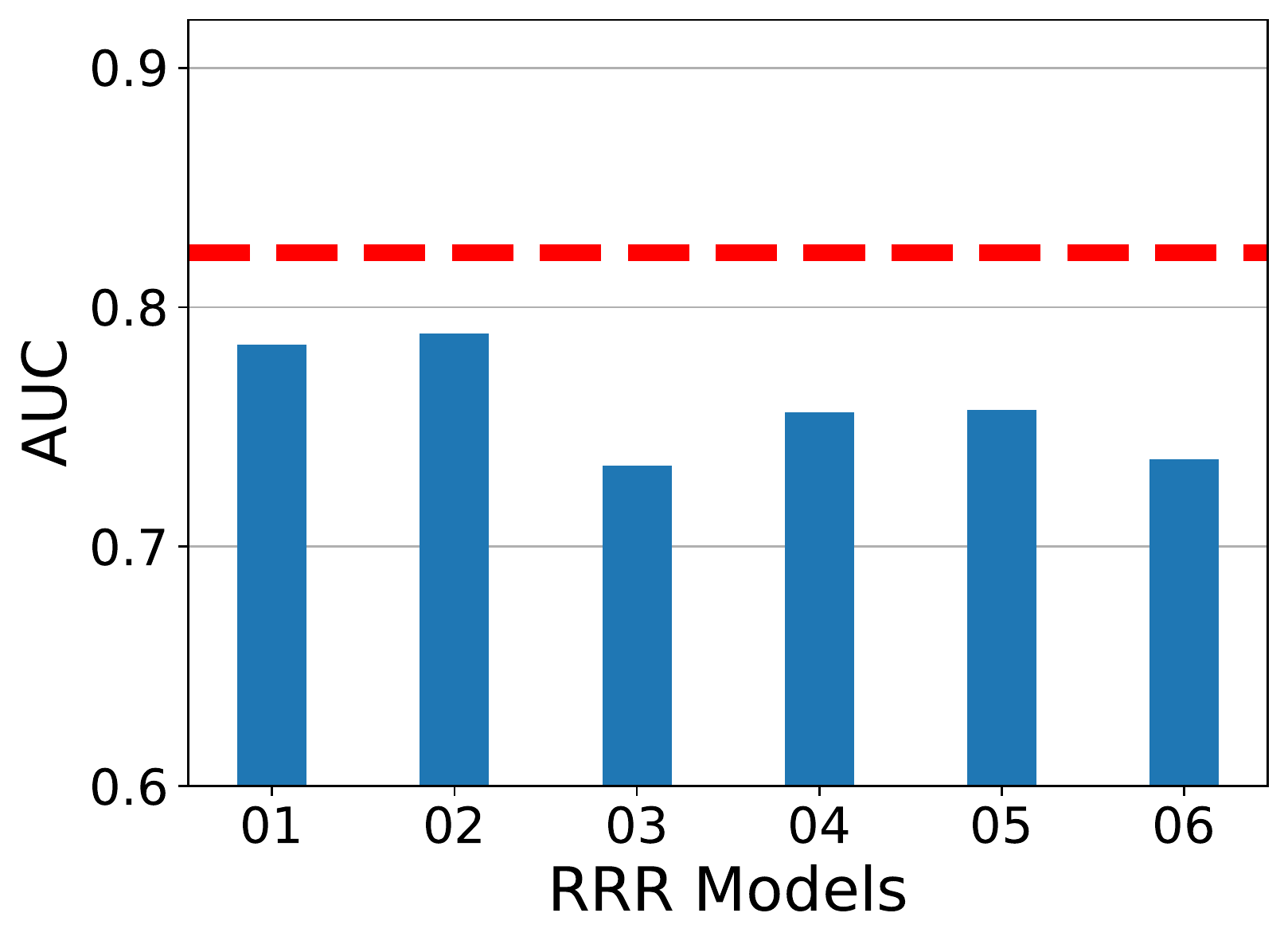}
   	\caption{Moving car model--parked car data \\ (2.4 GHz)}
   	\label{sf:mp-2.4}
\end{subfigure}
\begin{subfigure}[b]{0.375\textwidth}
	\centering
   	\includegraphics[width=\textwidth]{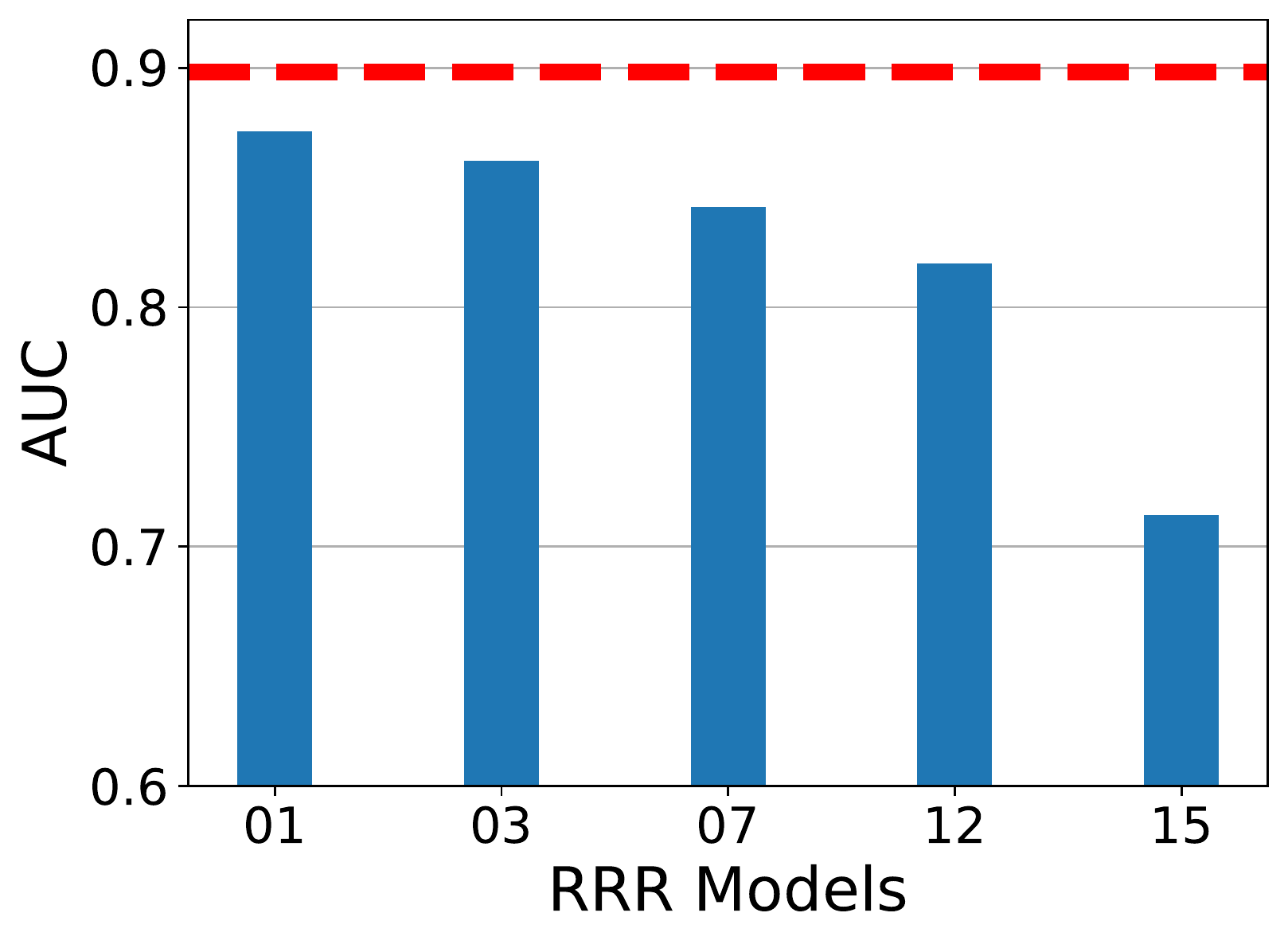}
   	\caption{Office model--parked car data \\ (5 GHz)}
    \label{sf:op-5}
\end{subfigure}
\caption{\gls{auc} of different \gls{rrr} models to mitigate attacks: (a) an adversary precollects \gls{csi} data in a similar car, (b) they train a neural network model on the rich office \gls{csi} and use it to classify copresence in a simpler environment of parked cars. The number of penalized features in the models increases from left to right; the dashed red line is the best achievable \gls{auc} without feature penalization; plot~\ref{sf:op-5} shows a subset of models.} 
\label{fig:rrr-adv}
\end{figure}

\autoref{sf:mp-2.4} and~\autoref{sf:op-5} depict the \gls{auc} performance of different \gls{rrr} models from~\autoref{fig:rrr-effect}.
The former, shows the models trained on the \textit{moving cars} \gls{csi} and tested on the \textit{parked car} data for 2.4 GHz, while the latter---the models trained on the \textit{office} \gls{csi} and tested on the \textit{parked cars} data for 5 GHz.
We see that the \gls{rrr} models with penalized features perform 12--25 \gls{auc} percentage points better when applied to the legitimate data compared to the adversarial data (cf.~\autoref{fig:rrr-effect} vs.~\autoref{fig:rrr-adv}).
In the adversarial case, the \gls{auc} drops from the model top-performing (cf. dotted red line in~\autoref{fig:rrr-adv}) more rapidly than in the legitimate case. 
Thus, the above attack can be mitigated by training a number of the \gls{rrr} models and combining their majority vote with thresholding for the final copresence decision (e.g., at least three models must have \gls{auc} above 0.9).

In the second attack, the adversary who knows that the neural network is trained on a single power level, which is typical for legitimate devices, increases the transmission power of their devices. 
We implement this attack by training our neural network on the \gls{csi} data combined from the \textit{office} and \textit{power} scenarios; in the latter scenario, we exclude high-power samples corresponding to non-copresent devices.
Thus, we evaluate the capability of our neural network to classify the unseen high-power \gls{csi} data. 

\autoref{tab:perf-adv} shows that under the power attack, the performance of \name drops by 16 and 11 \gls{auc} percentage points for 2.4 GHz and 5 GHz, respectively. 
We find that devices in adjacent Offices 1 and 2 and near open doors (cf.~\autoref{sf:setup-office}) are mostly affected by such an attack, accounting for the \gls{auc} drop.
The fact that only neighboring devices in such a challenging scenario (i.e., insufficient separation between Offices 1 and 2) are vulnerable to this attack suggests its limited scope. 
To mitigate the attack, we tried applying different \gls{rrr} models with penalized features, however, we observe a similar sharp drop in \gls{auc} from the top-performing model, as in~\autoref{fig:rrr-adv}. 
This indicates that the neural network simply lacks enough cues to distinguish copresent and non-copresent devices under such an attack. 
We discuss which other data, in addition to \gls{csi}, can improve the robustness of \name in~\autoref{sec:disc}. 
To alleviate the effect of the increased power attack, we find that including only 10\% of high-power samples brings classification performance very close to the baseline (cf. first row in~\autoref{tab:perf-adv}). 
This clearly helps improve the performance of the classifiers, as it provides more data close to the boundaries of the classes.
We highlight the utmost importance of training neural networks considering adversarial samples to harden \name.

\begin{table}
\small
\centering
	\caption{\gls{auc} performance of \name under the increased power attack for 2.4 GHz and 5 GHz bands.}
	\label{tab:perf-adv}
  \begin{tabular}{c|c|c}
  	\toprule
  	\multirow{2}{*}{\makecell{\gls{csi} data used for training \\ (included samples)}} & \multicolumn{2}{c}{Area Under the Curve (\gls{auc})} \\
  	& \xspace \xspace \xspace \xspace 2.4 GHz \xspace \xspace \xspace \xspace & 5 GHz \\
  	\midrule
  	Office + Power (all samples) & 0.9781 & 0.9967 \\
  	Office + Power (no high-power samples) & 0.8165 & 0.8825 \\ 
  	Office + Power (10\% of high-power samples) & 0.9313 & 0.9948 \\
  	\bottomrule
  \end{tabular}
\end{table}


\section{Discussion}
\label{sec:disc}
In this section, we provide relevant discussion points for \name.
\\
\textbf{Robustness.} 
Our evaluation shows that \name provides reliable copresence detection with error rates below 4\% and 1.5\% for 2.4 GHz and 5 GHz bands, respectively (cf.~\autoref{tab:perf-nadv}). 
We demonstrate the soundness of using \gls{csi} and neural networks for copresence detection (cf.~\autoref{sub:robust}), and find that \name can be hardened by training a number of neural network models, each relying on a different set of features, and combining their predictions (cf.~\autoref{subsec:adv}).
However, \name might be vulnerable to the increased power attack if the neural network is trained without considering the advanced capabilities of the adversary. 

To improve the robustness of \name, a number of physical layer metrics can be used in addition to \gls{csi}. 
For example, the \gls{snr} estimated directly in a wireless chipset should capture a unique noise pattern in the environment of copresent devices. 
Also, as shown by Won et al.~\cite{Won:2017}, \gls{csi} power levels are useful for traffic classification. In \name, these levels can be used to detect the increased power attack (cf.~\autoref{subsec:adv}). 
In addition, \glspl{pdp}, namely the squared magnitudes of the \gls{cir}, can be used as an additional input to a neural network, improving copresence detection accuracy by combining features from time (i.e., \gls{pdp} or \gls{cir}) and frequency (i.e., \gls{csi}) domains. 
Furthermore, devices running \name can measure the energy on different Wi-Fi channels to make copresence detection more reliable. 
Specifically, the energy sensed by copresent devices on a given Wi-Fi channel depends on its busyness in their environment (i.e., amount of traffic, other wireless devices using the channel in this environment). 
This energy will differ from measurements made by non-copresent devices, providing additional cues for differentiation. 
With several antennas, a direction from which a signal arrives can be determined, as we detail in the following discussion point.   
We also see that the temporal properties of \gls{csi} (i.e., behavior over time) increase the reliability of copresence detection (cf.~\autoref{subsec:non-adv}). 
\\
\textbf{Leveraging \gls{csi} Phase.}
Our results demonstrate that the \gls{csi} phase is less relevant for copresence detection than the magnitude, especially in complex environments with many obstacles and large amount of motion.
Prior research uses multiple antennas to remove a phase offset by finding the difference between phases received on different antennas, or obtain the relative phase by subtracting phases of two successively received frames~\cite{Ma:2019}.
The former approach only works if devices have multiple antennas, however, as we discuss in next point, it is unlikely that commodity \gls{iot} devices will receive them.  
The latter approach requires high packet rates (e.g., hundreds packets per second) to provide sufficient \gls{csi} granularity, while degrading fast in performance with reduced rates~\cite{Sen:2012, Shi:2017}. 
However, such high packet rates are impractical on \gls{iot} devices running on batteries.
The accurate phase estimation allows obtaining \gls{aoa} and \gls{tof}~\cite{Ma:2019}, which have the potential for increasing the robustness of \name to advanced attacks. 
\\
\textbf{Deployment Considerations.}
To use \name, devices performing copresence detection should be capable of extracting \gls{csi}. 
In recent years, several \gls{csi} extracting tools have emerged~\cite{Halperin:2011, Xie:2015, Nexmon:2017, Gringoli:2019}, enabling various devices such as routers, laptops, and smartphones with this capability. 
Utilizing these tools, security researchers discover severe vulnerabilities in proprietary firmwares of popular Wi-Fi chipsets~\cite{Classen:2019, Mantz:2019}, highlighting the necessity for open-source wireless stacks, which will provide better security and functionality (e.g., finer-grained control over a wireless chipset). 
Thus, we are positive that an increasing number of devices will receive the \gls{csi}-extraction capability, facilitating the deployment of \name. 

Another deployment consideration is using \name on devices that have a different number of antennas, namely single or multiple. 
As we discover in~\autoref{subsec:non-adv}, the \gls{csi} of one and two antenna devices varies, hindering the deployment of \name, if we cannot disable the second antenna and its spatial stream. 
This again urges the importance of finer-grained control over wireless chipsets.
For conventional 2.4 GHz and 5 GHz Wi-Fi, we do not expect low-power \gls{iot} devices (e.g., thermostat, smart lock) to have multiple antennas because of the size and energy constraints (i.e., two antennas require more processing), however, end-user devices such as smartphones, laptops, and routers will likely feature two or more antennas~\cite{Xue:2018}.
For higher frequency Wi-Fi based on mm-waves, even simple devices should be equipped with multiple antennas~\cite{Schultz:2013}.   
\\
\textbf{Copresence Distance.}
We consider devices to be copresent if they are located inside the same (office) room or car, which is commonly assumed by context-based copresence detection schemes~\cite{Karapanos:2015, Truong:2019, Fomichev:2019}. Our empirical results demonstrate that \name reliably detects copresence in rooms of up to $5\times6$ meter size and in typical passenger cars (e.g., hatchback). 
However, we see that the environments larger than $5\times6$ meter in size might become problematic for the 5 GHz band.
Also, the presence of prominent obstacles in the environment (e.g., sizeable wardrobe set, moving walls) can reduce the copresence detection accuracy of \name.
On the other hand, increasing the transmission power will extend the applicability range of \name. 
Overall, \name suits typical environments where copresence detection is applied, however, its parameters (e.g., used frequency band, transmission power) need to be adjusted for specific use cases. 


\glsresetall

\section{Conclusion}
\label{sec:concl}
In the age of the \gls{iot}, the demand for secure and usable authentication systems is on the rise. 
Context-based copresence detection enables such systems, allowing one device to verify proximity of another device based on their physical context (e.g., audio), eliminating user interaction. 
We propose \name, a robust context-based copresence detection scheme utilizing \gls{csi}.
As its main contribution, \name provides reliable copresence detection, including the challenging cases of low-entropy context (e.g., empty room with few events occurring) and insufficiently separated environments (e.g., adjacent rooms), and it does not require devices to have common sensors such as microphones. 
We implement and evaluate \name in five real-world scenarios, demonstrating its high classification accuracy in distinguishing copresent and non-copresent devices, the capability of working in real-time, and resilience to various attacks. 
The obtained error rates below 4\% show that \name outperforms the state-of-the-art copresence detection schemes, and it is deployable on off-the-shelf devices such as smartphones. 


\section{Acknowledgments}
\label{sec:ack}
We would like to thank Timm Lippert, Robin Klose, and Arash Asadi for their assistance in conducting this research.  
This work has been co-funded by the Research Council of Norway as part of the project Parrot (311197), the German Research Foundation (DFG) as part of projects A3 and B5G-Cell within the Collaborative Research Center (CRC) 1053 – MAKI, as well as the German Federal Ministry of Education and Research and the Hessian Ministry of Higher Education, Research, Science and the Arts within their joint support of the National Research Center for Applied Cybersecurity ATHENE.

\bibliographystyle{ACM-Reference-Format}
\bibliography{bibliography}

\section*{Appendix}
\appendix

\setcounter{footnote}{0}

\section{Background on the Right for the Right Reasons Method}
\label{sec:appx0}
Before describing the Right for the Right Reasons (\gls{rrr}\footnote{The original source code is available at \url{https://github.com/dtak/rrr}.}) method~\cite{Ross:2017}, we consider two topics: interpretability and hypotheses space.
Baehrens et al.~\cite{DBLP:journals/jmlr/BaehrensSHKHM10} points to an interesting fact about these network functions, specifically, the gradient of the network's output with respect to the input features $ \nabla_{\mathbf{x}_n} \hat{\mathbf{y}}_n $ is a vector normal to the decision boundary, and thus serves as a description of the model behavior near $ \mathbf{x}_n $. In the \gls{rrr} method~\cite{Ross:2017}, the authors propose to use these gradient vectors as explanations, and they further penalize those input gradients, making the network focus on relevant features and discarding irrelevant ones. With this approach, we can obtain explanations by finding which features are relevant for the prediction of an instance. 

For the penalization term, the authors introduce the annotation matrix $ \mathbf{A} \in \{0,1\}^{N \times D} $, which is a binary mask that indicates whether a feature should be relevant or not for a given instance. They then proceed to extend the standard loss functions by introducing a penalty $ \mathscr{P} (\mathbf{A}, \nabla_\mathbf{X} \hat{\mathbf{y}}) = \sum^N_{n = 1} \sum^D_{d = 1} \left( A_{n,d} \cdot \frac{\partial }{\partial x_{n,d}} \sum{}^Z_{z = 1} \log \left( \hat{y}_{n,z} \right) \right)^2 $ function on the input gradients controlled by a parameter $\lambda$, namely $ \widetilde{\mathscr{L}} (\boldsymbol{\theta}, \mathbf{X}, \mathbf{y}) = \mathscr{L} (\boldsymbol{\theta}, \mathbf{X}, \mathbf{y}) + \lambda \mathscr{P} (\mathbf{A}, \nabla_\mathbf{X} \hat{\mathbf{y}})$. This penalty function $ \mathscr{P} (\mathbf{A}, \nabla_\mathbf{X} \hat{\mathbf{y}}) $ and its influence value $\lambda$ guide the optimization algorithm to find optimal parameters given the restrictions imposed by $ \mathbf{A} $ on the features, while minimizing the prediction error. To understand the parameter $\lambda$, let us consider the two extremes: if $\lambda$ is low, the optimizer focuses only on the predictions, but if $\lambda$ is high, it will focus on the importance of the features and ignore the quality of the predictions. Here, we use the recommended $\lambda=1000$ because it keeps the values from the standard loss and the penalty on the same order of magnitude, as suggested by Ross et al.~\cite{Ross:2017}.
The above mentioned approach gives us interpretability by quantifying how much each feature contributes to the prediction of the network. However, it also gives us a way to obtain different classification hypotheses. 

Each classifier encodes one classification hypothesis, but there might be many different alternative explanations for the classification of a dataset. We can obtain different hypothesis by computing the input gradients to get a magnitude ratio, specifically, we divide the input gradients by the component with the maximum magnitude. We then compute the features per instance above a $c$ threshold, setting it to 0.67 according to the original work~\cite{Ross:2017}. After that, we aggregate the values for all the instances and remove the top most important features. This allows us to obtain different parameters for our neural network architecture that classify the data according to other alternative explanations, as they will not have access to the same input features.

\end{document}